\documentclass[aps,prx,twocolumn,superscriptaddress,nofootinbib,10pt]{revtex4-2}
\usepackage[utf8]{inputenc}
\usepackage[english]{babel}
\usepackage[T1]{fontenc}
\usepackage{graphicx} 
\usepackage{bm}
\usepackage{bbm}
\usepackage{amsfonts}
\usepackage{amsmath}
\usepackage{amsthm}
\usepackage{amssymb}
\usepackage[dvipsnames]{xcolor}
\usepackage{tikz}
\usepackage{hyperref}
\hypersetup{
  colorlinks  = true,
  urlcolor    = Blue,
  linkcolor   = BrickRed,
  citecolor   = PineGreen
}
\usepackage{lipsum}
\usepackage{comment}
\usepackage{tikz}
\usepackage[caption=false]{subfig}
\usepackage{booktabs}

\DeclareMathOperator{\Ex}{\mathbb{E}}
\DeclareMathOperator{\Var}{Var}
\DeclareMathOperator{\PP}{\mathbb{P}}
\DeclareMathOperator*{\argmax}{arg\,max}
\DeclareMathOperator*{\argmin}{arg\,min}

\newcommand*\diff{\mathop{}\!\mathrm{d}}
\newcommand{\cx}{\preceq_\text{cx}}
\newcommand{\degr}{\preceq_\text{deg}}

\begin{document}

\title{When the Noise Gets Noisy: Quantum Error Correction with Disordered Noise\texorpdfstring{\\}{ }and Misinformed Decoding}

\author{Cory T. Aitchison}
\thanks{These authors contributed equally; cta26@cam.ac.uk,\\lucas.english@sydney.edu.au.}
\affiliation{DAMTP, University of Cambridge, Wilberforce Road, Cambridge, CB3 0WA, UK}

\author{Lucas H. English}
\thanks{These authors contributed equally; cta26@cam.ac.uk,\\lucas.english@sydney.edu.au.}
\affiliation{School of Physics, University of Sydney, New South Wales 2006, Australia}

\author{Benjamin B\'{e}ri}
\affiliation{DAMTP, University of Cambridge, Wilberforce Road, Cambridge, CB3 0WA, UK}
\affiliation{TCM Group, Cavendish Laboratory, University of Cambridge, J. J. Thomson Avenue, Cambridge, CB3 0US, UK \looseness=-1}

\author{Dominic J. Williamson}
\affiliation{School of Physics, University of Sydney, New South Wales 2006, Australia}

\author{Campbell K. McLauchlan}
\affiliation{School of Physics, University of Sydney, New South Wales 2006, Australia}

\date{November 2026}

\begin{abstract}
Experimental demonstrations of quantum error correction often exhibit noise that varies across space and time, and attempting to characterize this disorder has been shown to substantially improve decoding performance. However, theoretical studies of quantum error correction typically assume fixed, perfectly known, homogeneous noise.
Here we model decoding with incomplete or inaccurate knowledge of disordered noise channels by treating physical error rates as random variables.
We show that the decoding performance is set by a small number of information-theoretic functionals of the noise distribution:
the entropy determines the threshold while Bhattacharyya coefficients bound the deep sub-threshold decay rate of logical failures.
Furthermore, our results give a quantitative measure of the benefits of characterizing this noise.
We numerically investigate both maximum-likelihood and minimum-weight-error decoding of surface codes, showing good agreement with theory across the code-capacity, phenomenological, and circuit-level noise settings, including with noise distributions seen in open-access superconducting devices. When it comes to error correction in realistic settings, reporting the average entropy and Bhattacharyya coefficient of noise distributions could therefore be more informative than the mean or median error rates of a device's components.
\end{abstract}
\maketitle

\section{Introduction}

Fault-tolerant quantum computation requires quantum error-correcting (QEC) codes, which encode logical quantum information into a subspace of a larger physical Hilbert space~\cite{Calderbank1996}.
The performance of QEC code families, including their thresholds, depends on the model of the noise which acts on the physical system~\cite{Aharonov1997,Knill1998,Dennis2002,Wan2026,English2026}. 
Theoretical QEC studies commonly assume noise channels with physical error rates that are uniform across the system, constant in time, and perfectly known by the decoder~\cite{Dennis2002,Wang2003}.
However, physical noise can exhibit spatial inhomogeneity, temporal variation, and correlations, including those arising from non-Markovian dynamics~\cite{Ng2009,McCutcheon2014,Figueroa-Romero2021,White2020,Biswas2025,Tan2026}
or from fabrication imperfections~\cite{Auger_2017,Strikis2023,Siegel_2023,Lin_2024,McLauchlan2024accommodating,Leroux_2025,Debroy_2025,Drouet2026,kim2026luciibmhardwareerror,mclauchlan2026errorcorrectedmemorylogicheavyhex,kim2026avoidingdefectsheavyhexagonarchitecture}. 
As quantum devices are scaled up to realize maximum utility with available hardware, the accuracy of QEC simulations based on these assumptions becomes increasingly important.

Recent literature has emphasized the value of characterizing the variability of noise in QEC demonstrations, including temporally drifting noise~\cite{Spitz_2018,Sriram2025,Fazio2026,Bhardwaj2026,Hoyt2026}, and spatial inhomogeneities~\cite{kim2026luciibmhardwareerror,Drouet2026,lee2026scalablequantumerrorcorrection,mclauchlan2026errorcorrectedmemorylogicheavyhex}. Improved characterization of the underlying noise model has been shown to improve decoding in both numerical simulations~\cite{Flammia_ACES,Hockings2025a,Hockings2025b,takou2025estimatingdecodinggraphshypergraphs,Hoyt2026} and experimental demonstrations~\cite{Chen_2022,Sivak_2024,Acharya2025,lee2026scalablequantumerrorcorrection,mclauchlan2026errorcorrectedmemorylogicheavyhex}. 
Understanding these improvements can help establish which variations in device performance can be accommodated through decoding and how accurately the noise must be characterized to achieve reliable logical operation.
This motivates the question as to what extent accurate characterization of disordered noise processes can improve QEC thresholds and sub-threshold scaling. This also presents the subtler problem of defining thresholds when the noise can no longer be described by a single physical error rate.

\begin{figure*}[t]
  \centering
  \subfloat[]{%
    \includegraphics[width=0.27\textwidth]{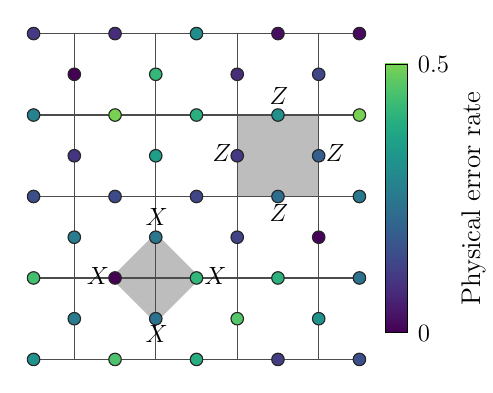}%
    \label{fig:errors_to_couplings}%
  }
  \hspace{5em}
    \subfloat[]{%
    \begin{minipage}[b]{0.5\textwidth}
        \centering
        \small
        \renewcommand{\arraystretch}{1.15}
        \setlength{\tabcolsep}{7pt}
        \begin{tabular*}{\linewidth}{@{\extracolsep{\fill}}lcc@{\hspace{14pt}}}
        \toprule
         & Threshold & Decay rate \\
        \midrule
        Characterized
         & \hyperref[sec:char-near]{$\overline{H}[P(p)]$}
         & \hyperref[sec:char-sub]{$\overline{B}[P(p)]$} \\
        Uncharacterized
         & \hyperref[sec:unchar-threshold]{$H(\overline{p})$}
         & \hyperref[sec:unchar-decay]{$B(\overline{p})$} \\
        Mischaracterized
         & \hyperref[sec:mis-near]{$\overline{H}[\Pi(p,\hat{p})]$}
         & \hyperref[sec:mis-sub]{$\overline{B}[\Pi(p,\hat{p})]$} \\
        \bottomrule
        \end{tabular*}
        \par\smallskip
        \begingroup
        \footnotesize\raggedright
        \textit{Threshold:} ML decoding; small correlations between $p$ allowed.\\
        \textit{Decay rate:} MWPM decoding on matchable codes; i.i.d.\ $p\ll 1$.
        \par\vspace{0.4cm}
        \endgroup
    \end{minipage}
    \label{fig:diagram}
  }

  \vspace{-1em}
    \subfloat[]{%
    \includegraphics[width=0.42\textwidth]{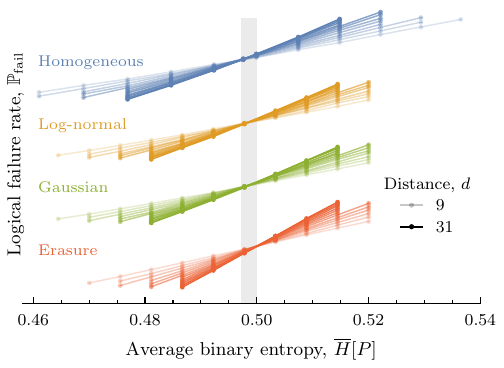}%
    \label{fig:intro-thresholds}%
  }
  \hspace{3em}
    \subfloat[]{%
    \includegraphics[width=0.45\textwidth]{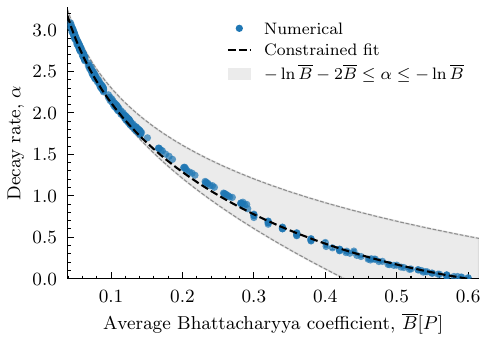}%
    \label{fig:intro-decay-rate}%
  }
\caption{
(a) Disordered noise channels are modeled with stochastic physical error rates (error parameters). Shown here is a surface code with qubits on edges. (b) Functionals governing the logical performance in different characterization settings. $P$ is the (marginal) probability density function for error parameters $p$ with a mean $\overline p$, and $\Pi$ is the joint distribution of true $p$ and decoder-estimate $\hat p$ error parameters. Bars denote distribution averages, $H$ denotes the binary entropy [Eq.~\eqref{eq:intro-entropy}] or cross entropy [Eq.~\eqref{eq:intro-bce}], and $B$ denotes the Bhattacharyya coefficient [Eq.~\eqref{eq:intro-bhat}] or its mischaracterized generalization [Eq.~\eqref{eq:intro-mis-bhat}]. (c) Characterized ML logical failure rates for surface codes with bit-flip noise following different error parameter distributions (vertical axes combined for illustration). All reach threshold at $\overline H[P] \approx 0.5$, marked by the gray region. (d) MWPM decay rates with characterized bit-flip noise following homogeneous, log-normal and Gaussian error parameter distributions. We show the predicted bounds, with the (leading-order) lower bound expected to hold at asymptotic $d$ in the deep-subthreshold regime (in the legend, we let $\overline{B}\equiv \overline{B}[P]$). We also fit to a curve $\alpha(\overline{B}[P])=-\ln\overline{B}[P]-2\overline{B}[P]-c\overline{B}[P]^{2}$, with $c \approx -2$ here.
}
\label{fig:combined_figure}
\end{figure*}

In this work, we present a framework for modeling disordered and mischaracterized noise, and show that the effects and benefits of noise characterization are governed by a small number of information-theoretic functionals. We demonstrate this numerically on surface codes across the code-capacity, phenomenological and circuit-level noise settings, with maximum likelihood (ML) and minimum weight perfect matching (MWPM) decoding~\cite{Cao2025,Higgott2025sparseblossom}. Figure~\ref{fig:combined_figure} provides an overview of the main concepts and results, which we outline in the section below.

\subsection{Main results}

We model disordered noise via local channels parameterized by random variables $p_i\sim P$, which set, for example, the bit-flip probability (with $i$ labeling a qubit at every step of a circuit). 
We first incorporate these error parameters into the statistical-mechanical mapping of QEC decoding~\cite{Dennis2002,Wang2003,ohnoPhase2004,Kovalev2015,Pryadko2020,Chubb2021, zhuNishimoris2023, chenNishimori2025, putzFlow2025a, Aitchison2026, mukherjeeNishimori2026}, whereupon replica trick arguments~\cite{Nishimori2002,Takeda2005} predict the threshold of the code under optimal, ML decoding. We show that this threshold, assuming perfect characterization of the noise---i.e.\ perfect knowledge of each realized $p_i$---is approximately set by the average entropy over the error parameter distribution $P$. 
In terms of the entropy of a binary noise channel,
\begin{equation}
    H(p) = -p \log p - (1-p)\log(1-p),
    \label{eq:intro-entropy}
\end{equation}
(with $\log$ denoting base $2$) threshold is reached when the noise distribution $P$ satisfies the condition
\begin{equation}
    \overline H[P] := \Ex_{p\sim P} \left[ H(p)\right] = H_c,
    \label{eq:intro-threshold}
\end{equation}
where $\sim$ indicates following the distribution of a random variable.
Here $H_{c}$ is a constant that is determined by the code and noise channels, and is equal to $\tfrac 12$ for codes which are self-dual under a Kramers-Wannier duality.
It is this functional of $P$, and not simply the mean error rate, that approximately sets the optimal threshold of the code. Different distributions $P$ for the same code and noise channels are all expected to attain threshold near this same critical $H_c$; we illustrate some examples for the surface code in Fig.~\ref{fig:intro-thresholds} where $H_c \approx \tfrac12$.
Using this framework, disorder is not necessarily detrimental to decoding: with perfect characterization, logical failures can be suppressed with increasing code distance (i.e.\ operating below threshold) even when the mean physical error rate is above the homogeneous noise threshold.

The behavior of QEC in different noise regimes is determined by different functionals of the noise distribution $P$. Below threshold, the average of the Bhattacharyya coefficient~\cite{Bhattacharyya1943} (for binary channels),
\begin{equation}
    B(p) = 2\sqrt{p(1-p)}
    \label{eq:intro-bhat}
\end{equation}
determines bounds on the decay rate~\cite{Bravyi2013,Acharya2025,Eickbusch2025,Atom2026} (with $\overline{\mathbb{P}}_\mathrm{fail}$ the logical failure rate averaged over both realizations of $p$ and faults\footnote{We also found our results to hold numerically for the average of $\ln\mathbb{P}_\mathrm{fail}$, which better captures typical values of the failure rate.})
\begin{equation}
    \alpha = - \lim_{d\rightarrow\infty} \frac1d \ln \overline \PP_\mathrm{fail},
\end{equation}
of perfectly characterized MWPM decoders; different noise distributions with the same average Bhattacharyya coefficient attain approximately equal decay rates, as illustrated in Fig.~\ref{fig:intro-decay-rate}. In the deep sub-threshold regime, we show that the decay rate follows a simple leading-order expression involving this quantity and predictable geometric constants $c_0$ and $c_1$,
\begin{equation}
    \alpha \geq -\ln \overline B[P] - \ln c_1 - c_0 \overline B[P] + O(\overline B[P]^2).
    \label{eq:intro-decay}
\end{equation}

We also study the more practical scenarios of uncharacterized decoders---where the distribution of error parameters is known, but not the individual realizations---and mischaracterized decoders---where an assumed error parameter realization differs from the realization that actually occurred---thus quantifying the costs and benefits of varying levels of characterization. Even with knowledge of the full distribution $P$, uncharacterized decoders can perform only as well as the decoder operating under homogeneous noise, i.e.\ only knowing the mean, thus losing all benefits associated with characterization.

For mischaracterized decoders, the threshold of the code is instead governed by the expected value of the cross entropy between the true $p$ and estimated $\hat p$ error parameters, which for binary channels is
\begin{equation}
    H(p,\hat p) = -p\log \hat p - (1-p)\log(1-\hat p).
    \label{eq:intro-bce}
\end{equation}
The decay rate of mischaracterized decoders is governed by a mismatched generalization of the Bhattacharyya coefficient with Chernoff tilt $s\geq 0$,
\begin{equation}
    B_s(p,\hat p) = (1-p) \left(\frac{\hat p}{1-\hat p}\right)^s + p\left(\frac{1-\hat p}{\hat p}\right)^s.
    \label{eq:intro-mis-bhat}
\end{equation}
The bounds of Eq.~\eqref{eq:intro-decay} then hold, with $\overline B[P]$ replaced by
\begin{equation}
    \overline B[\Pi] = \min_{s \geq 0} \Ex_{p,\hat p \sim \Pi}\left[ B_s(p,\hat p)\right].
\end{equation}

For both thresholds and decay rates, we show that under the simplifying model of $\ln p$ following a normal distribution---which can approximate realistic device noise distributions---attempting to characterize the noise is (to leading order) beneficial only when the variance of $\ln \hat p$ given $p$ is lower than the variance of $\ln p$ itself. 

Our work thus points to specific functionals---the average entropy, cross entropy, Bhattacharyya coefficient, and its mischaracterized generalization---as the relevant quantities that reveal most about the performance of realistic quantum devices.
Together, our results provide a framework for understanding noise in the more practical disordered setting where knowledge of noise parameters is limited, while quantifying the costs and benefits of characterizing such noise. 

\subsection{Layout of manuscript}
The rest of this manuscript is structured as follows. In Sec.~\ref{sec:notation}, we introduce the notation and conventions used in the following sections, describe the functionality of the optimal decoder under disordered noise and summarize the statistical mechanical mapping of QEC decoding. In Sec.~\ref{sec:characterized}, we quantify the performance of characterized decoders; in Sec.~\ref{sec:unchar} uncharacterized decoders; and in Sec.~\ref{sec:mischaracterized} mischaracterized decoders. Section~\ref{sec:numerics} presents numerical simulations of the preceding results.
Finally, we conclude in Sec.~\ref{sec:discussion} and outline future directions.

\section{Quantum error correction with disordered noise}\label{sec:notation}

In this section, we introduce our notation, models, and assumptions.
Section~\ref{sec:inhomogeneous-noise} introduces disordered noise channels and example error parameter distributions. Section~\ref{sec:decoders} discusses decoders, and their classification as characterized, uncharacterized, or mischaracterized. Section~\ref{sec:bg-sm} describes the statistical-mechanical mapping of decoding relevant for our main results.

\subsection{Disordered noise channels}
\label{sec:inhomogeneous-noise}

The analysis of the performance of QEC codes typically proceeds in three stages: (i) the noise processes affecting physical qubits and gates are approximated by simpler models (e.g. Pauli twirling, producing incoherent stochastic Pauli noise~\cite{Katabarwa2015}), (ii) error probabilities are assigned to these modeled processes, for example through device characterization~\cite{Harper2020,Hockings2025b} and (iii) simulation and analysis are carried out on the resulting channels. Noise processes, including stochastic Pauli errors and faulty measurements, can be described using the operator-sum representation~\cite{Kraus_1983,Beale2021}. For example, independent $X/Z$ noise takes the form
\begin{equation}
    \mathcal E_{p;Q}[\rho] = (1-p)\, \rho + p\,Q\rho Q, \quad Q= X,Z,
    \label{eq:noise-channel-ind}
\end{equation}
and for depolarizing noise,
\begin{equation}
    \mathcal E_{p,\text{dep}}[\rho] = (1-p)\rho + \frac{p}{3} \left( X\rho X + Y \rho Y + Z\rho Z\right),
    \label{eq:noise-channel-dep}
\end{equation}
where $p \in [0,1]$ is the physical error rate. More general noise channels can similarly be specified, including those involving correlations and multi-qubit Pauli errors.
In the phenomenological noise setting, data qubit errors are accompanied by measurement errors, which flip the signs of stabilizer measurement outcomes with some probability $p'$. We refer to each elementary stochastic error process as a mechanism, of which there are $n$ over the circuit or code.
We label possible outcomes of a mechanism as elements of an abelian group $G$: for a bit-flip channel this is $G =\{I,X\} \equiv \mathbb Z_2$, for depolarizing it is the (phaseless) Pauli group $\{I,X,Y,Z\} \equiv \mathbb Z_2^2$, and for measurement flips it is isomorphic to $\mathbb Z_2$. In the main text, we mostly consider binary mechanisms $|G|=2$ for simplicity, but generalizations for larger $G$ are discussed in Appendix~\ref{app:entropy-corr-noise}.

The analysis described above assumed that fixed parameters can characterize the physical noise of the system. However, more realistic models of noise, which account for environment-induced correlations, non-Markovianity, fabrication defects, and other physical effects, may cause spatiotemporal variation in the physical error rates over the duration of a device's operation and between shots~\cite{Auger_2017,Strikis2023,Siegel_2023,Lin_2024,McLauchlan2024accommodating,Leroux_2025,Debroy_2025,Drouet2026,kim2026luciibmhardwareerror,mclauchlan2026errorcorrectedmemorylogicheavyhex,kim2026avoidingdefectsheavyhexagonarchitecture,Spitz_2018,Sriram2025,Fazio2026,Bhardwaj2026,Hoyt2026,Tan2026}.
Moreover, any characterization of physical noise involves model approximations and estimation uncertainty, so decoders inevitably operate with potentially inaccurate priors. Their performance depends on how faithfully these priors capture the underlying noise processes.

In this manuscript, we address stages (ii) and (iii) to understand the effects on decoding performance of disordered, inhomogeneous noise and the uncertainty in its characterization. We consider a setting in which noise is applied through the following steps:
\begin{enumerate}
    \item every mechanism samples an error parameter $p$ with marginal probability density $P(p)$, independent and identically distributed (i.i.d.) across mechanisms unless otherwise stated;
    \item the corresponding qubits undergo noise channels, and measurement outcomes are probabilistically flipped, using the realized $p$ as parameter, and
    \item the decoder assumes an error parameter $\hat p$ for each mechanism, which may coincide with, or differ from, the true (realized) value $p$.
\end{enumerate}
We call the combination of steps 1.\ and 2.\ a \textit{disordered local noise channel}. We do not restrict to code-capacity noise here: a mechanism may be a noise channel on a qubit at a particular time in a circuit. To fix consistent signs of the couplings in the statistical-mechanical mapping, which ensures that the trivial outcome of each mechanism is the most likely, we assume that the parameter distribution has zero mass outside of the interval $[0,1-1/|G|]$\footnote{Any probability mass at $p=0$ can be treated by taking $0\ln0 = 0$; see also Ref.~\cite{Chubb2021}.}, i.e.,
\begin{equation}
    P(p) = 0, \quad \forall p \notin \left[0,1-\tfrac1{|G|}\right],\label{eq:P-mass}
\end{equation}
and that $P(p)$ is a valid probability density function,
\begin{equation}
    \int_0^{1-\frac1{|G|}}P(p)\ dp = 1.
\end{equation}
For the independent $X/Z$ noise channel in Eq.~\eqref{eq:noise-channel-ind} or for measurement flips, $|G|=2$; for the depolarizing channel of Eq.~\eqref{eq:noise-channel-dep}, $|G|=4$.

Generally, the random variables $\bm p = (p_1,\ldots,p_n)$ may be correlated across mechanisms, drawn from a joint distribution $\mathcal P(\bm p)$, which we assume has single-mechanism marginals all identically $P$; the threshold results of Secs.~\ref{sec:char-near}, \ref{sec:unchar-threshold}, and~\ref{sec:mis-near} hold at this level of generality. A physically motivated example of correlations, which we explore numerically, is the setting in which the $p$ are perfectly correlated in time. That is, we independently sample $p_i \sim P$ for each mechanism $i$ at the first timestep of a circuit, then reuse those values for each subsequent sampling of equivalent mechanisms at later times. (This is an example of a mixed memoryless (\textsf{MM}) quantum process~\cite{Taranto2024}, in that its process tensor is a classical mixture in $\bm p$ of memoryless quantum channels.)
Combining multiple mechanisms with non-identical marginal distributions, or of different mechanism types, remains a direction for future work. A more restrictive assumption is that the $p_i$ are i.i.d. $P$; this is used for the sub-threshold results of Secs.~\ref{sec:char-sub}, \ref{sec:unchar-decay}, and~\ref{sec:mis-sub}.

Our results hold for arbitrary probability distributions of the error parameters that satisfy the mass requirement Eq.~\eqref{eq:P-mass}. In this work, we focus on four key examples:
\begin{paragraph}{Homogeneous}
Using a Dirac delta function,
\begin{equation}
    P(p) = \delta(p-p^*), \quad p^* \in [0,\tfrac12],
\end{equation}
reproduces a binary Pauli channel with homogeneous error parameter $p^*$.
\end{paragraph}
\begin{paragraph}{Multimodal}
A mixture of two or more error parameters can be constructed as
\begin{equation}
    P(p) = \sum_{m=1}^M a_m \ \delta(p-p_m), \quad \sum_{m=1}^M a_m=1.
    \label{eq:bimodal}
\end{equation}
A special case of this is $M=2$, $p_1 = \tfrac12$, $p_2=0$, which reproduces the behavior of a simple erasure channel when used with a binary mechanism (and the decoder knows the locations at which $p=\tfrac12$)~\cite{Stace2009,Stace2010,Ohzeki2012}.
\end{paragraph}
\begin{paragraph}{Log-normal}
A physically motivated choice for $P$ is the log-normal distribution~\cite{Hockings2025a, Hockings2025b}, which resembles the noise distributions observed in some superconducting QEC demonstrations~\cite{Acharya2025} (although, see Appendix~\ref{app:CLN_sims} for examples of device noise that deviate from log-normality). The random variable
\begin{equation}p\sim\text{Lognormal}(\mu,\sigma^2)\end{equation}
is defined such that
\begin{equation}\label{eq:lognormal}
    \ln p \sim \text{Normal}(\mu,\sigma^2).
\end{equation}
The median of $p$ is $\exp\mu$, while its mean is
\begin{equation}
    \overline p = \Ex[p] = \exp\left(\mu + \frac{\sigma^2}{2}\right).
\end{equation}
\end{paragraph}
\begin{paragraph}{Gaussian}
The Gaussian noise channel draws physical error parameters from the random variable
\begin{equation}p\sim\mathrm{Normal}(\overline{p},\sigma^2).\end{equation}
\end{paragraph}
Because the log-normal has support on the domain $(0,\infty)$, and Gaussian on $(-\infty,\infty)$, we truncate both to satisfy the bounds in Eq.~\eqref{eq:P-mass}. We remove the outside probability mass and redistribute it proportionately to ensure normalization.\footnote{This shifts the mean and/or variance of $p$. In numerical analysis we treat this truncation exactly. For discussions, however, it is typically safe to ignore the effects of truncation provided the distribution's mass is concentrated sufficiently tightly.}
We discuss additional coding-theoretic properties of these disordered channels in Appendix~\ref{app:coding-theory}.

\subsection{Decoding quantum error correcting codes}
\label{sec:decoders}

In this section, we present a brief background on the decoding problem, and how we model different decoders with respect to disordered noise channels.

We assume here a code or circuit with binary fault mechanisms (independent noise channels and measurement errors); generalization to mechanisms with multiple nontrivial outcomes is discussed in Appendix~\ref{app:entropy-corr-noise}. Let $n$ denote the number of mechanisms, and encode their outcomes in the binary fault vector $\bm f = (f_1,f_2,\ldots,f_n) \in \mathbb Z_2^n$, where each entry indicates whether the corresponding noise channel produced a trivial ($0$) or nontrivial ($1$) error. Writing $\bm p = (p_1,\ldots,p_n)$ for the vector of realized per-mechanism error parameters, the probability of a fault vector is the product over mechanisms,
\begin{equation}
    \PP[\bm f;\;\bm p] = \prod_{i=1}^n p_i^{f_i} (1-p_i)^{1-f_i}.
    \label{eq:fault-prob}
\end{equation}
The stabilizer checks of the code are measured, producing a syndrome $\textsf s(\bm f) \in \mathbb Z_2^{m}$ that is trivial in the absence of faults, $\textsf s(\bm 0_n) = \bm 0_m$. The map from faults to syndromes is not injective, and the role of a decoder $\mathcal D$ is to take a syndrome and correct any accumulated logical error. It is sufficient to decode only to the correct equivalence class of faults---a set of faults with the same syndrome and same action on the encoded logical qubits. We denote a logical class by $\ell(\bm f)$.

The probability of a logical failure is then
\begin{equation}\label{eq:P_fail}
    \PP_\mathrm{fail}[\mathcal D;\bm p] = 1 - \sum_{\bm f} \PP[\bm f;\bm p] \ \delta_{\mathcal D[\textsf s(\bm f)] - \ell(\bm f)},
\end{equation}
with $\mathcal D[\textsf s(\bm f)]$ the output logical class from the decoder and $\delta_x$ the Kronecker delta.
This logical failure rate is minimized precisely when $\mathcal D[\textsf s]$ is the logical class of highest probability conditioned on each syndrome $\textsf s$. A decoder that does this is known as the maximum-likelihood (ML) decoder~\cite{Chubb2021}. Such decoders are difficult to implement in practice~\cite{Iyer2013}, as they require an enumeration (or a proxy thereof~\cite{Cao2025}) over all $\PP[\bm f;\bm p]$. A common, sub-optimal alternative is the minimum-weight-error decoder, which returns the logical class of the most probable fault~\cite{Iyer2013,Chubb2021} (equivalently, the minimum-weight fault, in the language of Sec.~\ref{sec:bg-sm}) consistent with the syndrome. 
If the decoding graph of the code is matchable, whereby every fault creates at most two syndrome flips, then this is efficiently computable using the minimum-weight perfect matching (MWPM) algorithm (here we use PyMatching~\cite{Higgott2025sparseblossom}).

The above formulation presumed nonrandom error parameters $\bm p$, which we now make stochastic. For a joint distribution $\mathcal P(\bm p)$ with marginals $P(p)$, the average logical failure rate is
\begin{equation}
    \overline{\PP}_\mathrm{fail}[\mathcal D;\mathcal P] = \Ex_{\bm p \sim \mathcal P}\left[\PP_\mathrm{fail}[\mathcal D;\bm p]\right].
\end{equation}
In the i.i.d.\ case $\mathcal P = P^{\otimes n}$ is the product measure, and we write $\overline{\PP}_\mathrm{fail}[\mathcal D;P]$ and $\Ex_{\bm p\sim P}$, omitting $\otimes n$ for simplicity.

If a decoder may access which $\bm p$ is realized in each shot (note this is not the same as knowing the true faults $\bm f$), then the minimizer of the average logical failure rate is just the ML decoder at each $\bm p$. We call this class of decoders \textit{characterized}, modeling the scenario where we are able to exactly characterize the noise, such as through perfect benchmarking of a device. This inclusion of $\bm p$ in the decoding process is an example of channel state information in coding theory~\cite{Elgamal2011}.

At the other extreme is a scenario where a decoder knows the distribution $\mathcal P(\bm p)$ but not the realized $\bm p$---we call these \textit{uncharacterized} decoders. This represents a known random device process, yet with unknown, uncharacterized samples. The noise channel that such decoders may access is
\begin{equation}
    \mathcal E_{\mathcal P}[\rho] = \int \mathcal P(\bm p) \mathcal E_{\bm p}[\rho] d\bm p 
\end{equation}
without additional channel state information. 
When the $p_i$ are independent, the marginal fault distribution $\Ex_{\bm p\sim P}[\PP[\bm f;\bm p]]$ factorizes exactly as $\prod_i \overline p^{f_i}(1-\overline p)^{1-f_i}$. From Eq.~\eqref{eq:P_fail}, we see that $\overline{\PP}_\text{fail}[\mathcal{D};\mathcal{P}]$ depends on $\mathcal{P}$ only through $\bar p$, so the optimal uncharacterized decoder is just the ML decoder for the homogeneous channel at $\overline p$, i.e.\ the ML decoder under $P(p)=\delta(p-\overline p)$ (see Appendix~\ref{app:bayes} for further details). An uncharacterized decoder is therefore equivalent to decoding while assuming that the underlying noise channels are homogeneous.

A third scenario is the most general: some information about the realized $\bm p$ is passed to the decoder, but this information is imperfect or uncertain. We model this with a joint distribution $\Pi(p_i, \, \hat{p}_i)$, where a decoder uses an assumed $\hat{\bm p}$ as its prior, and we call this \textit{mischaracterized} decoding. This represents realistic devices with imperfect attempts at characterization.

These decoders form the basis of our analytic and numerical studies in the remainder of this work, with Sec.~\ref{sec:characterized} focusing on characterized, Sec.~\ref{sec:unchar} on uncharacterized, and Sec.~\ref{sec:mischaracterized} on mischaracterized decoding.

\subsection{Statistical-mechanical mapping of decoding}
\label{sec:bg-sm}

The final ingredient needed for our analysis is the statistical-mechanical mapping for QEC codes~\cite{Dennis2002}. We present a brief description here for independent binary mechanisms, and refer the reader to other works for more detailed pedagogical discussions~\cite{Kovalev2015,Chubb2021, Pryadko2020, Wichette2025, English2025b, English2025, vennCoherentError2023, Behrends2025, behrendsSurface2025, Cao2026, Aitchison2026}.

A classical spin model is constructed such that each fault configuration is represented by a configuration of classical spin variables with suitably chosen interactions, and Boltzmann factor proportional to the probability of the fault. 
By choosing an appropriate Hamiltonian~\cite{Kovalev2015,Chubb2021}, summing over these Boltzmann factors to evaluate the partition function computes the probability of an entire logical class of faults.
Let $\mathcal G$ be the group of trivial faults---those with zero syndrome and no action on the logical qubits. These are stabilizers in stabilizer codes~\cite{Gottesman1997}, and gauge operators in subsystem codes~\cite{Poulin2005} and stabilizer circuits~\cite{Gidney2021, Delfosse2023, Aitchison2026}. Write the generators of $\mathcal G$ as binary vectors $\bm g_k \in \mathbb Z_2^n$, with $(\bm g_k)_i = 1$ if the fault includes mechanism $i$. Assign a classical Ising spin $S_k \in \{-1,1\}$ for each $\bm g_k$. For every fault mechanism $i=1,\ldots,n$ with error parameter $p_i$, we also assign a coupling strength (satisfying the Nishimori conditions of the resulting Hamiltonian)~\cite{Nishimori1981}
\begin{equation}
    K(p_i) = \frac12 \ln \left(\frac{1-p_i}{p_i}\right).
    \label{eq:nishimori}
\end{equation}
Additionally, given some reference fault $\bm f = (f_1,\ldots,f_n)$ (chosen to reproduce a given syndrome, for example, or drawn randomly for Monte Carlo sampling), we assign quenched coupling signs
\begin{equation}
    \eta(f_i) = \begin{cases}
        +1 & f_i = 0\\
        -1 & f_i=1.
    \end{cases}
\end{equation}
Putting these together, we obtain the Hamiltonian
\begin{equation}
    \mathcal H[\bm S,\bm f;\bm p] = -\sum_{i=1}^n \eta(f_i) K(p_i) \prod_{k}S_k^{(\bm g_k)_i},
\end{equation}
where $k$ runs over all spins. Here we consider binary mechanisms, $f_i\in\{0,1\}$. Mechanisms with more outcomes, such as the depolarizing channel of Eq.~\eqref{eq:noise-channel-dep}, also map to generalized Ising models, where multi-spin interactions couple the $X$ and $Z$ components of each fault~\cite{Chubb2021,Bombin2012}. We consider these in Appendix~\ref{app:entropy-corr-noise}. 

By construction, the Boltzmann factor for the Hamiltonian satisfies
\begin{equation}
    e^{-\mathcal H[\bm 1,\bm f;\bm p]} \propto \PP[\bm f;\bm p]
\end{equation}
where $\bm f$ is the reference fault, and the constant of proportionality is independent of $\bm f$ (but not of $\bm p$). Summing over all spin configurations $\bm S$ reproduces each fault $\bm f'$ with the same logical class and syndrome as $\bm f$. Summing the Boltzmann factor over all spin configurations is precisely the thermodynamic partition function, giving the relation:
\begin{align}
    \mathcal Z[\bm f;\bm p] &= \sum_{\bm S} e^{-\mathcal H[\bm S,\bm f;\bm p]}\notag \\ 
    &\propto \!\!\!\!\sum_{\bm f' \::\: \left\{\begin{matrix}\scriptstyle \mathsf s(\bm f') = \mathsf s(\bm f)\\\scriptstyle \ell(\bm f') = \ell(\bm f)\end{matrix}\right.}\!\!\!\!\PP[\bm f';\bm p].
    \label{eq:partition}
\end{align}

The partition function depends on $\bm f$ only through its syndrome and logical class, so we write $\mathcal Z_{\ell,\bm{\mathsf s}}[\bm p] := \mathcal Z[\bm f;\bm p]$ for any representative fault $\bm f$ with $\mathsf s(\bm f)=\bm{\mathsf s}$, $\ell(\bm f)=\ell$. Different logical classes are constructed using representative faults from their class.
An ML decoder, which we denote $\mathcal D_\mathrm{ML}^{\bm p}$, chooses the logical class with the largest partition function among those consistent with the syndrome. The logical failure rate is then the probability that a fault was drawn from a different logical class, averaged over possible syndromes:
\begin{equation}
    \PP_\mathrm{fail}\left[\mathcal D_\mathrm{ML}^{\bm p};\bm p\right] = \sum_{\bm{\mathsf s}} \PP[\bm{\mathsf s};\bm p]\left(1-\frac{\max_\ell \mathcal Z_{\ell,\bm{\mathsf s}}[\bm p]}{\sum_{\ell}\mathcal Z_{\ell,\bm{\mathsf s}}[\bm p]}\right),
\end{equation}
where $\PP[\bm{\mathsf s};\bm p]$ is the probability of the syndrome $\mathsf s$. 

The ML threshold of the code corresponds to an order-disorder phase transition of the statistical mechanical model to which the code maps~\cite{Dennis2002}. Ratios of partition functions probe the free energy cost of inserting defect lines along the support of logical operators; in the ordered phase, these free energy costs diverge in the thermodynamic limit, and the code is below threshold.

The prototypical example of this mapping is the toric code under independent bit-flip code-capacity noise, which maps to the random-bond Ising model (RBIM) on the two-dimensional square lattice~\cite{Dennis2002}. Spins are assigned to vertices $i$, with couplings along the edges $\langle i,j\rangle$. The Hamiltonian is
\begin{equation}
    \mathcal H[\bm S,\bm f;\bm p] = -\sum_{\langle i,j\rangle} \eta(f_{ij}) K(p_{ij}) \ S_i S_j.
\end{equation}
For homogeneous noise, this model undergoes an order-disorder phase transition at $p_c =0.1092212(4)$~\cite{Wan2026}.

\section{Characterized decoders}
\label{sec:characterized}

In this section, we investigate the behavior of characterized decoders; Sec.~\ref{sec:char-near} explains the threshold result from Eq.~\eqref{eq:intro-threshold}, while Sec.~\ref{sec:char-sub} bounds their decay rates as in Eq.~\eqref{eq:intro-decay}. The ideas developed here will later be generalized to uncharacterized and mischaracterized decoders in Sec.~\ref{sec:unchar} and~\ref{sec:mischaracterized} respectively.

\subsection{Characterized thresholds}
\label{sec:char-near}

By applying the results of \textcite{Nishimori2002} and \textcite{Takeda2005} to statistical mechanical models with disordered couplings, we discuss the threshold of characterized decoders, with a full derivation given in Appendix~\ref{app:entropy}. We show that the critical point is set by the average binary entropy of its (marginal) parameter distribution, $P$.

\subsubsection{Critical average binary entropy}
\label{sec:char-near-crit}

Recall from Sec.~\ref{sec:bg-sm} that the ML decoding threshold of a code coincides with the order-disorder phase transition of an associated classical spin model. For disordered noise, the coupling strengths $K(p_i)$ are random, with $\bm p \sim \mathcal P$; the phase diagram of the spin model is thus determined by a distribution rather than a single number, and the critical point becomes a critical manifold in the space of parameter distributions. Since the optimal characterized decoder is the ML decoder at each realized $\bm p$, this manifold is precisely the threshold of optimal characterized decoders. We first focus on disordered noise with binary mechanisms, $|G|=2$, for simplicity; mechanisms with multiple outcomes (e.g. depolarizing noise) are discussed at the end of this section.

For a homogeneous $p$, \textcite{Nishimori2002} and \textcite{Takeda2005} conjectured via a replica and duality argument that the critical point of a model, $p_c$, and of the model placed on its dual lattice, $p_c^*$, satisfy $H(p_c) + H(p_c^*) = 1$, where
\begin{equation}
    H(p) := -p\log p - (1-p)\log(1-p)
    \label{eq:binary-entropy}
\end{equation}
is the binary entropy, using the convention $\log \equiv \log_2$. Applying these arguments to the statistical mechanical models from QEC decoding with disordered noise, we find a condition for critical marginal distributions (see Appendix~\ref{app:entropy} for details). Specifically, write the average binary entropy as
\begin{equation}
    \overline H[P] := \Ex_{p\sim P}[H(p)],
\end{equation}
and let $P^*$ denote some marginal distribution on the dual lattice. Then, two critical marginals $P_c, P_c^*$ satisfy
\begin{equation}
    \overline H[P_c] + \overline H[P_c^*] = 1.
    \label{eq:critical-H}
\end{equation}
This constrains $H_c$ in relation to the dual model: for self-dual models, such as the square-lattice RBIM of Sec.~\ref{sec:bg-sm} and many common QEC codes~\cite{English2026}, $\overline H[P_c^*] = \overline H[P_c]$ and thus they all should satisfy $H_c = \tfrac12$.

\begin{figure}
    \centering
    \includegraphics[width=.8\linewidth]{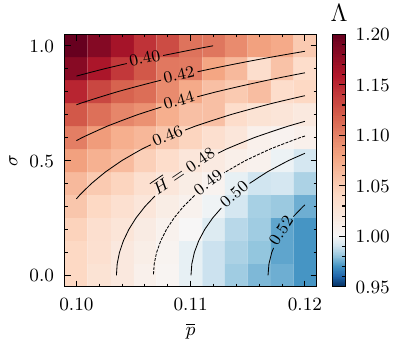}
    \caption{Ratio of logical failure rates $\Lambda = \overline{\PP}_\mathrm{fail}(d=11)/\overline{\PP}_\mathrm{fail}(d=13)$ for characterized ML decoding of surface codes with bit-flip noise, following log-normal error parameters with specified $\overline{p}$ and $\sigma$. $\Lambda$ is a proxy for threshold: when $\color{Maroon}\Lambda > 1$ the system is below threshold; $\color{MidnightBlue}\Lambda<1$ is above. Contours show average binary entropy $\overline H[P]$, with threshold ($\Lambda = 1$) forming a critical manifold $0.495\lesssim \overline{H}[P] \lesssim0.5$.
    See Appendix~\ref{app:sup-numerics} for simulation details.
    }
    \label{fig:char_contour}
\end{figure}

For non-self-dual models, this predicts a critical manifold that is a level set of $\overline H$ at $H_c = 1- \overline{H}[P_c^*]$.
For a given code (or circuit) and noise channels, if distribution $P_1$ is critical with $\overline H[P_1] = H_c$, then any distribution $P_2$ with $\overline H[P_2] = H_c$ is also predicted to be critical, regardless of the microscopic details of the distribution (including correlations with sufficiently small correlation lengths over the error parameters). The relevant quantity for analyzing and comparing characterized decoder performance near threshold is thus the average binary entropy: parameter distributions that share the same mean $\overline p$ but differ in average binary entropy can be at different positions relative to threshold. For example, Fig.~\ref{fig:char_contour} shows ML decoding of surface codes with bit-flip noise and $p_i \overset{\text{i.i.d.}}{\sim}\mathrm{Lognormal}$ while varying $\overline p$ and $\sigma$. Threshold is reached (ratio $\Lambda = \overline{\PP}_\mathrm{fail}(d)/\overline{\PP}_\mathrm{fail}(d+2)=1$) along the critical manifold $\overline H[P] \approx \tfrac12$.

The derivation of Eq.~\eqref{eq:critical-H} uses the replica trick, taking $\frac1N \ln \Ex[\mathcal Z^N]$ as $N\to0$.\footnote{This can also be understood as an $m\to1$ replica limit, where $m=N+1$ is the exponent of $\mathcal Z$ after expanding the expectation to include the probability of disorder realizations (faults) on the Nishimori line.} To leading order in $N$, the resulting fixed-point condition depends only on the marginal $P$, assuming the probability of correlations in $\mathcal P$ spanning regions that contain logical operators is exponentially suppressed (see Appendix~\ref{app:entropy-corr} and Ref.~\cite{Sriram2025}). The mixed memoryless processes with error parameters fixed in time (but not in space) satisfy this assumption.

\begin{table*}[t]
    \caption{Comparison of bit-flip and erasure thresholds, which Eq.~\eqref{eq:critical-H} predicts to satisfy $H(p_c) = q_c$. Here $p_c$ is the critical point of the RBIM on the given lattice (the ML bit-flip threshold of the corresponding toric code), and $q_c$ the erasure threshold from percolation on the dual lattice.}
    \label{tab:bitflip-erasure}
    \setlength{\tabcolsep}{12pt}
    \begin{tabular*}{\linewidth}{lcccc}
        \toprule
        Lattice & $p_c$ & $H(p_c)$ & $q_c$ & $|H(p_c)-q_c|/H(p_c)$ \\
        \midrule
        Square ($2$D) & $0.1092212(4)$~\cite{Wan2026} & $0.4975624(12)$ & $\tfrac12$~\cite{Stace2009} & $0.5\%$ \\
        Honeycomb ($2$D) & $0.0675(5)$~\cite{deQueiroz2006} & $0.3565(19)$ & $2\sin(\pi/18)\approx0.347$~\cite{Sykes1964} & $3\%$ \\
        Cubic ($3$D) & $0.2327(3)$~\cite{Ozeki1998} & $0.7827(5)$ & $0.75118818(10)$~\cite{Aizenman1983,Wang2013} & $4\%$ \\
        \bottomrule
    \end{tabular*}
\end{table*}

\subsubsection{Multi-outcome fault mechanisms}

The same duality arguments extend to mechanisms with more than two outcomes, such as the depolarizing channel of Eq.~\eqref{eq:noise-channel-dep} (see Appendix~\ref{app:entropy-corr-noise} for additional details). Recall that we denote $G$ as the group of a mechanism's outcomes ($|G|=4$ for single-qubit Paulis). In this setting, rather than evaluating the binary entropy, we evaluate the Shannon entropy,
\begin{equation}
    H_G(p) = -\sum_{g \in G} \PP[g;p] \log \PP[g;p],
    \label{eq:Shannon}
\end{equation}
where $\PP[g;p]$ is the probability of that mechanism outcome under the noise channel parameterized by $p$. 
For depolarizing mechanisms, we can relate the Shannon entropy to the binary entropy through $\overline H_G[P] = \overline H[P] + \overline p\log(|G|-1)$.
Denoting $\overline H_G[P]$ as the average Shannon entropy of its fault distribution, Eq.~\eqref{eq:critical-H} generalizes to define the critical marginals $P_{c}$ and $P_{c}^{*}$ through
\begin{equation}
    \overline H_G[P_c] + \overline H_G[P_c^*] = \log|G|.
    \label{eq:critical-H-group}
\end{equation}

This criterion provides a unified framework for predicting thresholds in a range of contexts, many of which are well-studied in the literature. As the toric code under homogeneous single-qubit depolarizing noise is self-dual, Eq.~\eqref{eq:critical-H-group} predicts $H(p_c) + p_c\log3 = 1$. This expression is satisfied by the zero-rate hashing bound $p_c \simeq 0.18929$~\cite{Bennett1996}, which can be compared with the value obtained via Monte Carlo simulations for the toric code, $p_c \simeq 0.189(3)$~\cite{Bombin2012}. Another context covered by Eq.~\eqref{eq:critical-H-group} is the erasure channel: modeling an erasure as a fully depolarizing channel ($p=\tfrac34$), we reproduce the duality estimate of Ref.~\cite{Ohzeki2012} for the toric code with qubit loss. Equation~\eqref{eq:critical-H-group} does not, however, determine how mechanisms of different types combine, as in circuit-level noise: the fixed-point condition is a system-wide average, which cannot capture the behavior of an arbitrary system.

\subsubsection{Relationship between thresholds}

We emphasize that the results presented above rely on the conjectured fixed-point condition of \textcite{Nishimori2002} and on the validity of the replica continuation across the considered parameterizations, and so should be regarded as an approximate predictor of the critical manifold. Numerical evidence nevertheless points to this approximation being accurate across a range of models. The Nishimori--Nemoto condition $H(p_c)=\tfrac12$~\cite{Nishimori2002} is known to be close to, but not exactly, the true critical point of the self-dual square-lattice RBIM: numerics give a threshold $p_c\approx0.1092212(4)$, equivalently $H(p_c)\approx0.4975624(12)$~\cite{Wan2026}, within $\approx0.5\%$ of $\tfrac12$. 

Let us now consider two very different families of error parameter distributions on the same lattice. We first represent a bit-flip channel through the homogeneous point mass $P_{\text{hom},p_X}(p) = \delta(p-p_X)$,
with $\overline H[P_{\text{hom},p_X}] = H(p_X)$. We next represent the erasure limit of the bimodal distribution of Eq.~\eqref{eq:bimodal}, $P_{\text{er},q}(p) = (1-q)\delta(p) + q\,\delta(p-\tfrac12)$, which we call a (binary) \textit{erasure channel} with erasure rate $q$. This distribution has $\overline H[P_{\text{er},q}] = q$.
Equation~\eqref{eq:critical-H} predicts that the bit-flip threshold $p_c$ and erasure threshold $q_c$ both lie on the critical manifold $H(p_c) = q_c = H_c$.

We compare the thresholds of toric codes on different lattices in Table~\ref{tab:bitflip-erasure}. Erasure thresholds can often be determined analytically: a logical failure becomes likely once the erased ($p=\tfrac12$, $K=0$) bonds percolate along the support of a logical operator, which lives on the dual of the spin-model lattice. The free-energy cost of inserting the logical operator then vanishes, and two partition functions that differ in equivalence class by that logical operator become equal. For $2$D toric codes, the logical operators are strings~\cite{Bombín_2014}, and $q_c$ is the bond-percolation threshold of the dual lattice. 
The square lattice is self-dual, and the dual of the honeycomb lattice is the triangular lattice.
In $3$D, the toric code's $X$-error threshold lies at the critical point of the $3$D RBIM on the cubic lattice~\cite{Xu2026}; its logical operators are membranes, so erasures cause failure once the erased bonds form a spanning membrane on the dual lattice (plaquette percolation), which occurs at $q_c = 1 - p_\mathrm{bond}$, where $p_\mathrm{bond} =0.24881182(10)$ is the cubic-lattice bond-percolation threshold~\cite{Aizenman1983,Wang2013}. As shown in the table, the predicted relationship between the bit-flip and erasure thresholds agrees to within $0.5\%$ (square), $3\%$ (honeycomb) and $4\%$ (cubic). 

The relation $\overline{H}[P_c] = q_c$ allows us to estimate a mixed-noise phase boundary from the pure-erasure threshold $q_c$, which is often easier to compute via the percolation transition. For example, consider the bimodal channel
\begin{equation}
    P(p) = (1-q)\delta(p-a q) + q\delta\!\left(p-\tfrac12\right).
\end{equation}
$q$ is the erasure probability, and each unerased qubit undergoes a bit flip with probability $p_X=aq$, with $0\leq aq\leq\tfrac12$. For a fixed noise ratio $a$, the predicted critical erasure rate $q_a$ satisfies
\begin{equation}
    q_a + (1-q_a)H(aq_a) = q_c.
\end{equation}
Varying $a$ then traces the predicted bit-flip--erasure phase boundary, including for codes that are not self-dual. We discuss a self-dual example in Appendix~\ref{app:erasure}.

These comparisons have considered only homogeneous or bimodal i.i.d.\ error parameters and code-capacity noise; we use numerics to include phenomenological and circuit-level noise, depolarizing noise, time-correlated error parameters, and other distributions in Sec.~\ref{sec:char-sim}.

\subsection{Characterized decay rates}
\label{sec:char-sub}

Utility-scale quantum computing will likely necessitate physical error rates far below threshold. In this regime, the relevant figure of merit is how rapidly the logical failure rate is suppressed as the code distance grows~\cite{Acharya2025}. 
At large $d$, $\overline{\PP}_\mathrm{fail} \approx e^{-\alpha d}$, and we quantify this suppression by the asymptotic decay rate~\cite{Bravyi2013}
\begin{equation}
    \alpha[\mathcal D; P] := \lim_{d\to\infty} -\frac1d \ln\overline{\PP}_\mathrm{fail}[\mathcal D; P],
    \label{eq:decay-rate}
\end{equation}
where $\overline{\PP}_\mathrm{fail}[\mathcal D;P]$ is the parameter-averaged logical failure rate. At finite $d$, $\alpha$ can be estimated via linear regression using a code family with varying distances. Here $\alpha$ is related to the error suppression factor $\Lambda := \overline{\PP}_\mathrm{fail}(d)/\overline{\PP}_\mathrm{fail}(d+2)$ reported in experiments~\cite{Acharya2025,Eickbusch2025,Atom2026} by $\Lambda \approx e^{2\alpha}$.

Here we focus on MWPM decoding of topological codes (circuits) subject to noise channels producing matchable syndrome graphs (detector error models); this includes the surface or toric codes. 
We show that the decay rates of characterized decoders are determined not by the average binary entropy, but instead by the average Bhattacharyya coefficient defined through~\cite{Richardson2008}
\begin{equation}
    \overline B[P] := \Ex_{p\sim P}[B(p)], \qquad B(p) := 2\sqrt{p(1-p)}.
    \label{eq:bhat}
\end{equation}
We assume here that the $p_i \overset{\text{i.i.d.}}{\sim} P$, and faults are from uncorrelated binary channels; relaxing these assumptions is left to future work.

We also show that in the deep sub-threshold regime, path-counting arguments predict the asymptotic form
\begin{equation}
    \alpha[\mathcal D; P] \geq -\ln \overline B[P] - \ln c_1 - c_0 \overline B[P] + O\big(\overline B[P]^2\big),
    \label{eq:char-decay}
\end{equation}
where $c_0$ and $c_1$ are constants set by the geometry of the decoding graph. The analytic approximation requires $\overline B[P] \ll 1$, which is a natural regime for the study of QEC. For example, a homogeneous error parameter $p = 10^{-3}$ gives $B(p) \approx 0.06$. Correlations between $p_i$ prevent exchanging the order of expectation and product over fault locations (a key step in our approach), and we thus leave their generalization to future work. While these results are derived under the assumption of MWPM decoding, the decay rates of MWPM and ML decoding have been found numerically to often (but not always) coincide at low physical error rates~\cite{English2025, zhaoDivergence2026}.

\subsubsection{Average Bhattacharyya coefficient}
\label{sec:char-sub-rates}

Here we show that the characterized MWPM logical failure rates of matchable codes, and thus their decay rates, are bounded by functions of the average Bhattacharyya coefficient given in Eq.~\eqref{eq:bhat}.

We first consider an upper bound on $\overline{\PP}_{\mathrm{fail}}$. Assume some fault $\bm f$ occurs. Let $\mathcal D$ be the characterized MWPM decoder, which, given the syndrome $\mathsf s(\bm f)$ and the realized parameters $\bm p$, returns the logical class of the most probable consistent fault $\bm f^\star$.\footnote{On a matchable decoding graph this is implemented by minimum-weight perfect matching with edge weights $2K(p_i)$~\cite{Higgott2025sparseblossom}.} Suppose the decoder fails, $\ell(\bm f^\star) \neq \ell(\bm f)$. Then $\bm f \oplus \bm f^\star$ has trivial syndrome and nontrivial logical action, and so contains a simple self-avoiding path $\bm C$ that is homologically nontrivial (e.g., a path between two boundaries on a planar surface code) and is a nontrivial logical operator. $\bm f \oplus \bm f^\star$ may also contain faults with trivial syndrome and logical action, i.e.\ contractible cycles. Since $\bm f$ and $\bm f^\star$ differ at every location of $\bm C$, and $\bm f^\star \oplus \bm C$ is no more probable than $\bm f^\star$ by optimality, the likelihood ratio satisfies the inequality
\begin{equation}
    X_{\bm C}(\bm f;\bm p) := \frac{\PP[\bm f \oplus \bm C;\bm p]}{\PP[\bm f;\bm p]} = \frac{\PP[\bm f^\star;\bm p]}{\PP[\bm f^\star \oplus \bm C;\bm p]} \geq 1,
\end{equation}
where, from Eq.~\eqref{eq:fault-prob},
\begin{equation}
    X_{\bm C}(\bm f;\bm p) = \prod_{i \in \bm C} \left(\frac{p_i}{1-p_i}\right)^{1-2f_i}.
    \label{eq:likelihood-ratio}
\end{equation}
That is, the set of all faults $\bm f$ (given a $\bm p$) that produce logical failures is contained within the set of faults satisfying the likelihood ratio inequality for at least one $\bm C$, 
\begin{equation}
    \{\text{failing } \bm f;\bm p\} \subseteq \bigcup_{\bm C} \{X_{\bm C}(\bm f;\bm p)\geq 1\},
    \label{eq:union-fail}
\end{equation}
where the union runs over all simple paths that are nontrivial logical operators.
This is generally not an equality as ties ($X_{\bm C} = 1$) may be broken probabilistically by the decoder. Also, a fault with $X_{\bm C}(\bm f;\bm p)\geq1$ may be decoded correctly if another path $\bm C'$ exists that is logically trivial, but satisfies $\PP[\bm f \oplus \bm C'] \geq \PP[\bm f \oplus \bm C]$. This may occur when there are more than two boundaries, for example. We also impose here that the error parameters are strictly $0 < p_i < \tfrac12$ so that the likelihood ratio is informative. Erasure channels are therefore not covered by the predictions in this section.

The union bound over Eq.~\eqref{eq:union-fail} then gives
\begin{equation}
    \PP_\mathrm{fail}[\mathcal D;\bm p] \leq \sum_{\bm C} \PP_{\bm f}\big[X_{\bm C}(\bm f; \bm p) \geq 1\big],
\end{equation}
where $\PP_{\bm f}$ denotes the probability over faults at a fixed $\bm p$. Equality in the union bound holds only when possible events are disjoint; $\bm C$ that share support have strongly correlated events, and so this inequality is not expected to be tight in general.

This tail probability of $X_{\bm C}$ has no clean form for general $\bm C$ and $\bm p$, and so we replace it with an expectation using the Chernoff-Cram\'er bound in $\ln X_{\bm C}$~\cite{Dembo1998, Boucheron2013},
\begin{equation}
    \label{eq:chernoff}
    \PP_{\bm f}[\ln X_{\bm C} \geq 0] \leq \Ex_{\bm f}\left[ e^{s\ln X_{\bm C}}\right] = \Ex_{\bm f}\left[X_{\bm C}^s\right],
\end{equation}
for any arbitrary $s > 0$. Since the faults are independent, the expectation factorizes:
\begin{equation}
    \Ex_{\bm f}\left[X_{\bm C}^s\right] = \prod_{i \in \bm C} \left[(1-p_i)^{1-s} p_i^{s} + p_i^{1-s} (1-p_i)^{s}\right].
\end{equation}
The terms in this product are known as $s$-skewed Bhattacharyya affinity coefficients and appear also in classical coding-theoretic Chernoff bounds~\cite{Bhattacharyya1943, nielsenGeneralized2014, nielsenRevisiting2022}.
Each factor is convex in $s$ and symmetric under $s \leftrightarrow 1-s$, and thus minimized (giving the tightest bound) at $s = \tfrac12$, where it equals the normal Bhattacharyya coefficient $B(p_i)$ shown in Eq.~\eqref{eq:bhat}. Combining this result with the union bound and averaging over $\bm p \sim P$, using the independence of the $p_i$ to move the expectation through the product,
\begin{align}
    \overline{\PP}_\mathrm{fail}[\mathcal D;P] &\leq \Ex_{\bm p \sim P} \left[ \sum_{\bm C} \prod_{i\in \bm C} B(p_i) \right ] \notag \\ 
    &= \sum_{\bm C} \prod_{i\in \bm C} \overline B[P] \notag \\
    &= \sum_{\lambda \geq d} N(\lambda)\, \overline B[P]^{\lambda}
    \label{eq:pfail-bhat}
\end{align}
where we sum over paths of length $\lambda$ with multiplicity $N(\lambda)$, and $\lambda \geq d$ by the definition of the code distance. When the noise is homogeneous, this reduces to Eq.~(22) of \textcite{Beverland2019}.

If the number of possible starting points grows polynomially with $d$, $N(\lambda)$ is at most a polynomial in $d$ times the number of self-avoiding walks of length $\lambda$ on the underlying lattice, which grows as $\kappa^{\lambda + o(\lambda)}$. This $\kappa$ is the lattice's growth constant ($\kappa \approx 2.638$ for the square lattice in $2D$~\cite{Jacobsen2016} and $\kappa\approx4.684$ for the cubic lattice in $3D$~\cite{Clisby_2013}). Equation~\eqref{eq:pfail-bhat} vanishes exponentially in $d$ whenever $\kappa\, \overline B[P] < 1$ (this inequality reduces to the bound of \textcite{Dennis2002} for homogeneous noise).

We now consider a lower bound on $\overline{\PP}_\mathrm{fail}$. Choose a nontrivial logical operator with path $\bm C_0$ of length $d$. The map $\bm f \mapsto \bm f \oplus \bm C_0$ partitions all faults into disjoint pairs $\{\bm f, \bm f\oplus\bm C_0\}$ with the same syndrome but different logical classes. Any decoder must fail on at least one of them; the pair of faults contributes at least the smaller of their two probabilities to the failure rate. Summing over all $\bm f$ counts each such pair twice, giving the lower bound
\begin{align}
    \PP_\mathrm{fail}[\mathcal D;\bm p] &\geq \sum_{\{\bm f,\, \bm f\oplus\bm C_0\}} \min\big(\PP[\bm f;\bm p],\, \PP[\bm f\oplus\bm C_0;\bm p]\big) \notag\\
    &= \tfrac12 \sum_{\bm f} \PP[\bm f; \bm p] \ \min(1, X_{\bm C_0}) \notag \\
    &= \tfrac12\, \Ex_{\bm f}\big[\min(1, X_{\bm C_0})\big] \notag \\ 
    &\geq \tfrac12\, \PP_{\bm f}\big[X_{\bm C_0} \geq 1\big].
\end{align}
This is the usual lower bound on the error of distinguishing two hypotheses~\cite{LeCam1973}, i.e.\ faults $\bm f$ and $\bm f\oplus\bm C_0$.
Jointly over $p_i$ and $f_i$, $\ln X_{\bm C_0}$ is a sum of $d$ i.i.d.\ terms. Cram\'er's theorem~\cite{Dembo1998} then states that the Chernoff bound of Eq.~\eqref{eq:chernoff} is exponentially tight once averaged over $\bm p$, $\Ex_{\bm p}\big[\PP_{\bm f}[X_{\bm C_0} \geq 1]\big] = \overline B[P]^{d}\, e^{o(d)}$, and so
\begin{equation}
    \overline{\PP}_\mathrm{fail}[\mathcal D;P] \geq \tfrac12\, \overline B[P]^{d}\, e^{o(d)}.
    \label{eq:pfail-lower}
\end{equation}

Together, Eqs.~\eqref{eq:pfail-bhat} and~\eqref{eq:pfail-lower} thus bound $\alpha$ between functions of $\overline B[P]$ alone. Taking $-\tfrac1d\ln$ and letting $d\to\infty$, the polynomial and sub-exponential prefactors vanish, and whenever $\kappa\,\overline B[P] < 1$,
\begin{equation}
    -\ln\overline B[P] - \ln\kappa \;\leq\; \alpha[\mathcal D;P] \;\leq\; -\ln\overline B[P].
    \label{eq:alpha-bracket-kappa}
\end{equation}
This leading term, $-\ln \overline{B}[P]$, is the (averaged) Bhattacharyya distance~\cite{Bhattacharyya1943} or Chernoff information~\cite{nielsenRevisiting2022}. In the deep sub-threshold regime, we show next that these bounds can be made tighter.

\subsubsection{Deep sub-threshold decay rates}
\label{sec:char-sub-path}
We bound the path sum of Eq.~\eqref{eq:pfail-bhat} by counting a larger class of walks, whose weighted sum can be evaluated explicitly. This gives a Taylor series in $\overline B[P]$, which we truncate in the deep sub-threshold regime.

Assume that the decoding graph is embedded in a lattice so that each nontrivial logical path $\bm C$ connects the layers $x=0$ and $x=d$ for some lattice coordinate $x$. Every simple logical path is then a walk on the infinite extension of this lattice, with the restriction that it never immediately reverses its preceding step (backtracks)~\cite{fitznerNonbacktracking2013}. For a fixed starting point at $x=0$, let $W_x$ be the sum of weights over all such walks ending in layer $x$, with each walk of length $\lambda$ assigned weight $\overline B[P]^\lambda$. The number $M(d)$ of possible starting points grows at most polynomially with $d$. Counting walks from all these starting points gives $\sum_{\lambda \geq d} N(\lambda)\overline B[P]^\lambda \leq M(d)W_d$, since the infinite-lattice count also includes walks that leave the original finite patch.

To derive $W_x$, we consider a recursive expression for how many ways there are of ending a valid walk at coordinate $x$. We first count the number of steps a walk may take. Let there be $c_0$ steps (an edge on the lattice) that take a walk from $x \mapsto x$, $c_{\pm 1}$ from $x\pm 1\mapsto x$, $c_{\pm 2}$ from $x\pm 2\mapsto x$, and so on. Each such step multiplies the weight by $\overline B[P]$. In the unrotated surface code, the decoding graph is a square lattice and we have $c_0 = 2$ (if $x$ traverses the horizontal lattice direction, then there are two steps that keep the walk within column $x$: up or down), $c_{\pm 1} = 1$ (left or right), and $c_w =0$ for $|w| \geq 2$. In the rotated surface code, there are no steps that keep a walk within the same column, $c_0 = 0$, but $c_{\pm1}=2$ ways of traversing from $x\pm1$.

Some of these steps immediately backtrack the preceding step, which is disallowed. Subtracting these pairs of forbidden steps leaves an admissible walk ending at the same coordinate $x$. Given a nonempty walk ending at $x$, there are $v - 1$ such forbidden pairs, where $v = \sum_{w \in \mathbb Z}c_w$ is the valency of the lattice vertex: the first step of the pair can follow any edge (of which there are $v$) except for the previous step in the walk. Each pair contributes a factor of $\overline B[P]^2$ for the two steps. 

Together, these arguments give the recurrence relation (with $\overline B \equiv \overline B[P]$ for notational simplicity)
\begin{equation}
    W_x = \overline B \sum_{w \in \mathbb Z} c_w W_{x+w} -\overline B^2 \left(v - 1\right)W_x,\quad x>0.
    \label{eq:path-count}
\end{equation}
Such recurrences admit geometric solutions $W_x =W_0 \omega^x$ for some $\omega$. For simplicity, we restrict now to lattices that are local, with $c_w = 0$ for $|w| \geq 2$, and symmetric along $x$, with $c_1 = c_{-1}$. This gives the quadratic in $\omega$
\begin{equation}
    \begin{split}
        c_1 \overline B \omega^2 -b(\overline B)&\omega + c_1 \overline B = 0, \\ 
        b(\overline B) = 1-c_0\overline B + &[c_0+2c_1-1]\overline B^2.
    \end{split}
\end{equation}
Taking the smaller root, such that $W_x \rightarrow 0$ as $x \rightarrow \infty$ (for $\overline B[P] < 1/(v-1)$ the total weight is finite), we get 
\begin{equation}
    \omega = \frac{b(\overline B) - \sqrt{b(\overline B)^2-4c_1^2\overline B^2}}{2c_1\overline{B}}.
\end{equation}
We thus have that $\overline{\PP}_{\mathrm{fail}}[\mathcal D; P] \leq M(d)W_0 \omega^d$. Taking $-\tfrac1d\ln$ as $d\rightarrow \infty$ and expanding about $\overline B[P] = 0$ gives
\begin{equation}
    \alpha[\mathcal D; P] \geq -\ln \overline{B}[P] - \ln c_1 - c_0 \overline B[P] + O\left(\overline B[P]^2\right).
    \label{eq:alpha-gtrsim}
\end{equation}
For the unrotated surface code, for example, we get 
\begin{equation}
    \alpha[\mathcal D;P] \geq - \ln \overline B[P] - 2 \overline B[P] - \tfrac23 \overline B[P]^3 + O\left(\overline B[P]^4\right).
\end{equation}
Equation~\eqref{eq:alpha-gtrsim} thus tightens the lower bound of Eq.~\eqref{eq:alpha-bracket-kappa}, giving
\begin{equation}
    -\ln\overline B[P] - \ln c_1 - O\big(\overline B[P]\big) \;\leq\; \alpha \;\leq\; -\ln\overline B[P].
    \label{eq:alpha-bracket}
\end{equation}
At finite $d$, the bounds gain $O(\ln d /d )$ terms from the polynomial prefactors. With the unrotated surface code, $c_1 = 1$ and the bounds agree at leading order, and coincide as $\overline B[P]\rightarrow 0$.

In general, these bounds are not exact predictors of $\alpha$. The expansion of $\ln\omega$ gives the leading small-$\overline B[P]$ behavior, which worsens as $P$ approaches threshold. The union bound in Eq.~\eqref{eq:pfail-bhat} also over-counts failure events for all codes. We consequently regard these bounds as guides for predicting the form of the true decoder exponent, which can be computed and compared numerically. We have assumed throughout this section that all $p_i$ are drawn i.i.d.; with correlated $p_i$ the $\overline B[P]^{\lambda}$ would be replaced by $\Ex_{\bm p\sim\mathcal P}\big[\prod_{i\in\bm C}B(p_i)\big]$. We leave investigations of such correlations to future work.

\section{Uncharacterized decoders}
\label{sec:unchar}
We now compare characterized decoders with uncharacterized decoders, which lack information about the realized $p_i$ and must decode based on the distribution $P$ alone (in the i.i.d.\ case this reduces to decoding over a homogeneous channel). The ideas here build directly upon the theory developed in the previous section.

\subsection{Uncharacterized thresholds}
\label{sec:unchar-threshold}

Equation~\eqref{eq:critical-H} identified $\overline H[P] = H_c$ as the threshold condition for characterized decoders, with (in the i.i.d.\ case) an uncharacterized decoder reaching threshold when $H(\overline p) = H_c$. What does this difference tell us about the gains available from characterization? First note that $\overline H[P]$ has a direct information-theoretic interpretation: without knowledge of $p_i$---i.e.\ for uncharacterized decoders---a fault $f_i$ is a Bernoulli random variable with probability $\overline p = \Ex_{p\sim P}[p]$, and entropy $H(f_i) = H(\overline p)$. Knowing $p_i$, its conditional entropy is instead $H(f_i\,|\,p_i) = \Ex_{p\sim P}[H(p)] = \overline H[P]$. Characterizing $p_i$ therefore supplies, per mechanism, the mutual information
\begin{equation}
    \Delta H[P] := I(f_i;p_i) = H(\overline p) - \overline H[P]
    \label{eq:delta-H}
\end{equation}
between the fault and its error parameter~\cite{Richardson2008}. This is non-negative by Jensen's inequality, as the concavity of $H$ gives for any distribution $P$ with mean $\overline p$,
\begin{equation}
    \overline H[P] \;\leq\; H(\overline p),
    \label{eq:jensen}
\end{equation}
with equality if and only if $P$ is a homogeneous point mass (in which case $p_i$ carries no additional information about $f_i$). In other words, for fixed mean $\overline p$, the distribution with the highest average binary entropy is the homogeneous delta function (see Appendix~\ref{app:coding-theory}).

Holding $\overline p$ fixed while spreading probability mass across fault mechanisms thus lowers $\overline H[P]$, potentially improving the performance of a characterized decoder, while these benefits are lost on uncharacterized decoders. For example, consider a distribution $P$ with mean $\overline p$ such that $H(\overline p) = H_c$ sits at the critical point.
For i.i.d.\ $p_i$, an optimal uncharacterized decoder is equivalent to the ML decoder for the homogeneous channel $p = \overline p$, so the uncharacterized decoder is operating at threshold.
However, $\overline H[P] = H_c - \Delta H[P] < H_c$, and so a properly characterized ML decoder is below threshold and exhibits a lower logical failure rate.

What level of noise disorder is required to attain practical benefit?
Characterized error parameter distributions in superconducting devices are sometimes found to resemble log-normal distributions~\cite{Acharya2025, Hockings2025a}, for which $\overline H[P]$ decreases with $\sigma$ at fixed $\overline p$. For $p\sim\mathrm{Lognormal}(\mu,\sigma^2)$, using $\Ex[p\log p] = \overline p(\log\overline{p} + \sigma^2/[2\ln 2])$ gives 
\begin{equation}
    \Delta H[P] \approx \overline p\frac{\sigma^2}{2\ln 2},
    \label{eq:delta-H-lognormal}
\end{equation}
assuming $p \ll 1$, i.e.\ the contributing mass in $P$ is distributed $\ll 1$.
To give a sense of scale, suppose we have a log-normal error parameter distribution with $\overline p = 0.1$ and $\sigma^2 = 0.1$ (typical values of $p$ falling approximately between $0.06$ and $0.15$). Characterization provides $\Delta H[P] \approx 0.007$ bits of additional information compared to an uncharacterized decoder. If we vary the mean $\bar p$, the surface code bit-flip threshold is then reached at $\overline p \approx 0.112$, compared with $0.109$ when uncharacterized (or if the noise is perfectly homogeneous).

When the $p_i$ are correlated, Eq.~\eqref{eq:jensen} and the interpretation of $\Delta H$ as a per-mechanism mutual information are unchanged, since both depend only on the marginal $P$. For uncharacterized decoding, however, the homogeneous decoder at $\overline p$ is no longer the optimal uncharacterized decoder as the marginal fault probabilities do not factorize.
Nevertheless, if we treat this as a mischaracterized decoder with $\hat p_i = \overline p$, as we shall explore in Sec.~\ref{sec:mis-near}, the threshold of an uncharacterized homogeneous-weight decoder is predicted at
\begin{equation}
    H(\overline p) = H_c.
\end{equation}

\subsection{Uncharacterized decay rates}
\label{sec:unchar-decay}

The uncharacterized MWPM decoder (which uses uniform edge weights) has its asymptotic decay rate bounded by Eq.~\eqref{eq:char-decay} with $\overline B[P]$ replaced by $B(\overline p)$. This follows from the fact that with i.i.d.\ $p_i$, the marginal fault distributions are identical to those drawn from a homogeneous $\overline p$. Thus, an uncharacterized decoder's average logical failure rate equals its failure rate under homogeneous noise at $\overline p$. Since $B$ is strictly concave, Jensen's inequality again gives
\begin{equation}
    \overline B[P] \leq B(\overline p),
    \label{eq:jensen-B}
\end{equation}
with equality if and only if $P$ is a homogeneous point mass. In the deep sub-threshold regime, the resulting improvement to the decay rate from characterization is
\begin{align}
    \Delta\alpha[P] &:= \alpha[\mathcal D_\mathrm{char}; P] - \alpha[\mathcal D_\mathrm{unchar}; P] \nonumber\\
    &\approx \ln \left(\frac{B(\overline p)}{\overline B[P]}\right) + c_0 \big(B(\overline p) - \overline B[P]\big),
    \label{eq:delta-alpha}
\end{align}
which follows whenever $\alpha$ saturates the lower bound of Eq.~\eqref{eq:char-decay}. $\Delta\alpha$ is non-negative to this order, and vanishes only for homogeneous noise. 

An alternative comparison is to consider the ratio of the asymptotic form of the logical failure rates,
\begin{equation}
    \frac{\overline{\PP}_\mathrm{fail}[\mathcal D_\mathrm{char};P] }{ \overline{\PP}_\mathrm{fail}[\mathcal D_\mathrm{unchar};P]} \approx e^{-\Delta\alpha\, d} \approx \left[\frac{\overline B[P]}{B(\overline p)} \right]^d,
\end{equation}
or in terms of the error suppression factor, 
\begin{equation}
    \frac{\Lambda_\mathrm{char}}{\Lambda_\mathrm{unchar}} \approx e^{2\Delta\alpha} \approx \left[\frac{B(\overline p)}{\overline B[P]}\right]^2.
\end{equation}

For the log-normal distribution of Eq.~\eqref{eq:lognormal} concentrated at small $p\ll 1$, $B(p) \approx 2\sqrt p$. Using $\Ex[\sqrt p] = e^{\mu/2 + \sigma^2/8}$ and $\sqrt{\overline p} = e^{\mu/2 + \sigma^2/4}$, the leading term of Eq.~\eqref{eq:delta-alpha} becomes
\begin{equation}
    \Delta\alpha \approx \frac{\sigma^2}{8} \quad \Longrightarrow \quad \frac{\Lambda_\mathrm{char}}{\Lambda_\mathrm{unchar}} \approx e^{\sigma^2/4} \approx 1 + \frac{\sigma^2}{4},
    \label{eq:delta-alpha-lognormal}
\end{equation}
independent of $\mu$ (neglecting truncation). In other words, a characterized decoder below threshold which experiences log-normal noise with mean $\overline p$ performs like a decoder facing homogeneous noise at the effective error parameter $p_\mathrm{eff}$ defined through $B(p_\mathrm{eff}) = \overline B[P]$. Sufficiently far below threshold, $p_\mathrm{eff} \approx (\Ex[\sqrt p])^2 = \overline p\, e^{-\sigma^2/4} < \overline p$. For example, log-normal noise with $\sigma^2 = 0.1$ and $\overline p = 0.01$ gives the characterized decoder a $2.5\%$ higher $\Lambda$. Equivalently, it performs decoding at an effective physical error rate that is $2.5\%$ smaller ($p_\mathrm{eff}\approx 0.00975$).

\section{Mischaracterized decoders}
\label{sec:mischaracterized}

In this section, we examine how accurate characterization must be to improve decoding performance. We model this via the mischaracterized decoder, which assumes parameters $\hat{\bm p}$ that may differ from the true $\bm p$, which together follow the joint probability distribution $\Pi(p_i, \hat p_i)$. This requires generalizations to the functionals discussed earlier in Sec.~\ref{sec:characterized}.

\subsection{Mischaracterized thresholds}
\label{sec:mis-near}

The mischaracterized ML decoder selects the logical class with the largest partition function $\mathcal Z_{\ell,\bm{\mathsf s}}[\hat{\bm p}]$, evaluated at the assumed parameters $\hat{\bm p}$ and not the true parameters $\bm p$. In the corresponding spin model, the coupling strengths are set by the assumed parameters, $K(\hat p_i)$, while the signs $\eta(f_i)$ are still drawn with the true parameters $p_i$. Whenever $\hat p_i \neq p_i$, the model therefore lies off the Nishimori line.\footnote{See also the discussions of decoding off the Nishimori line in Ref.~\cite{Wichette2025}.} As with Sec.~\ref{sec:char-near}, we allow for correlations between the $p_i$ (and now $\hat p_i$) at different fault locations $i$, as long as large correlated regions along the support of logical operators are sufficiently suppressed, and we assume the marginal distributions at every $i$ to be identically $\Pi(p_i,\hat p_i)$.

In generalizing the replica calculation of Appendix~\ref{app:entropy} for mischaracterization, we instead utilize the average binary cross entropy, 
\begin{align}
    \overline H[\Pi] &:= \Ex_{(p,\hat p) \sim \Pi}\left[H(p,\hat p) \right],\\
    H(p, \hat p) &:= -p \log \hat p - (1-p) \log(1-\hat p).
    \label{eq:bce}
\end{align}
The fixed-point condition for a model and its dual is then given by
\begin{equation}
    \overline H[\Pi] + \overline H[\Pi^*] = 1,
    \label{eq:critical-mischar}
\end{equation}
where $\Pi^*$ is the joint parameter distribution of the dual model. The binary cross entropy exceeds the binary entropy by the Kullback--Leibler (KL) divergence, 
\begin{equation}
    H(p,\hat p) - H(p) = D(p\,\|\,\hat p),
\end{equation}
where
\begin{equation}
    D(p\,\|\,\hat p) := p \log \left(\frac{p}{\hat p}\right) + (1-p) \log \left(\frac{1-p}{1-\hat p}\right) \geq 0.
\end{equation}
Averaging over $\Pi$,
\begin{align}
    \overline H[\Pi] &= \overline H[P] + \overline D[\Pi],\\
    \overline D[\Pi] &:= \Ex_{(p,\hat p)\sim\Pi}\left[D(p\,\|\,\hat p)\right],
    \label{eq:bce-kl}
\end{align}
where $P$ is the marginal distribution of $p$. 

In this notation, $\overline D[\Pi]$ is a measure of mischaracterization, and it plays the same role as $\Delta H$ in Sec.~\ref{sec:unchar-threshold}. For a characterized decoder, $\hat p = p$, so $\overline D[\Pi]=0$ and Eq.~\eqref{eq:critical-mischar} reduces to Eq.~\eqref{eq:critical-H}. For an uncharacterized decoder, $\hat p = \overline p$ gives $\overline H[\Pi] = H(\overline p)$ exactly. 
More generally, if $H_c$ is the critical average binary entropy of a model, a mischaracterized decoder is predicted to reach threshold when
\begin{equation}
\overline H[P] + \overline D[\Pi] = H_c.
\end{equation}
Mischaracterization thus acts as $\overline D[\Pi]$ bits of additional entropy that lowers the threshold.

For arbitrary $\Pi$, Eq.~\eqref{eq:critical-mischar} is less justified than the critical condition of Eq.~\eqref{eq:critical-H}. The fixed-point conjecture of Refs.~\cite{Nishimori2002,Takeda2005} is justified by the multicritical point lying on the Nishimori line; off-Nishimori, it is less concrete. The naive replica trick predicts a phase boundary $p_{c}\to0$ as $T\to 0$~\cite{Ohzeki2008a, Ohzeki2008b, Ohzeki2009}, whereas the critical probability in the RBIM has been found numerically to be $p_{c}\gtrsim 0.103$ for all $T$ below the Nishimori line (equivalently, $\hat p < p$)~\cite{Wang2003,Amoruso2004,Merz2002,Thomas2011}. 
Near the line, 
\begin{equation}
    D(p\,\|\,\hat p) = (\hat p - p)^2/[2p(1-p)\ln 2] + O\big((\hat p-p)^3\big)
\end{equation}
is quadratic in $\hat p - p$, so Eq.~\eqref{eq:critical-mischar} correctly predicts that the threshold is maximal on the Nishimori line ($\hat p= p$). Above the Nishimori temperature ($\hat p > p$), numerical comparisons to the predicted phase boundary show modest agreement even at larger mismatch~\cite{English2026}. We therefore treat Eq.~\eqref{eq:critical-mischar} as a leading-order prediction for weak mischaracterization.

As an example, suppose each estimate carries an independent multiplicative log-normal error, $\hat p_i = p_i z_i$ with $\ln z_i \sim \mathrm{Normal}(-\hat\sigma^2/2, \hat\sigma^2)$. Here, the $-\hat\sigma^2/2$ offset preserves the mean $\Ex[\hat p_i|p_i] = p_i$ and $\hat\sigma^2$ sets the size of the error.\footnote{The estimates should also be truncated as in Eq.~\eqref{eq:P-mass}; we neglect this in our discussions here, which is justified when $\hat\sigma \ll -\ln(2p_i)$. Numerical simulations use truncation.} The first logarithm in $D(p_i\,\|\,\hat p_i)$ averages exactly, $\Ex[\log \hat p_i\,|\,p_i] = \log p_i - \hat\sigma^2/(2\ln 2)$, since $\ln z_i$ is normal. The second requires an expansion about $p_i$ to second order. Using $\Var[\hat p_i\,|\,p_i] = p_i^2(e^{\hat\sigma^2}-1) \approx p_i^2\hat\sigma^2$, we find
\begin{equation}
    \Ex[\log(1 - \hat p_i)\,|\, p_i] \approx \log(1-p_i) - \frac{\hat \sigma^2}{2\ln 2}\,\frac{p_i^2}{(1-p_i)^2},
\end{equation}
for $\hat\sigma \ll 1$. Combining the two terms gives us
\begin{equation}
    \overline D[\Pi] \approx \frac{\hat\sigma^2}{2\ln2}\,\Ex_{p\sim P}\left[\frac{p}{1-p}\right].
    \label{eq:kl-lognormal}
\end{equation}
For $p \ll 1$, $\overline D[\Pi]$ is thus proportional to the mean $\overline p$.

For a sense of scale, take $\hat\sigma^2 = 0.1$. For homogeneous noise on the square-lattice RBIM, this gives $\overline D[\Pi] \approx 0.009$ bits at threshold, which lowers the predicted threshold from $p_c \approx 0.109$ to $p_c \approx 0.106$. A smaller mischaracterization, $\hat\sigma^2 = 0.01$, gives $\overline D[\Pi] \approx 0.0009$ bits and $p_c \approx 0.1089$.

Observe that this cost has the same form as the benefits from characterization in Eq.~\eqref{eq:delta-H-lognormal}, with $\hat\sigma^2$ in place of $\sigma^2$. To leading order, characterization therefore improves the threshold only if the logarithmic error of the estimates, $\hat\sigma^2$, is smaller than the variation in the true parameters, $\sigma^2$. We find the same criterion for below threshold, as we shall discuss now.

\subsection{Mischaracterized decay rates}
\label{sec:mis-sub}

We now extend the arguments in Sec.~\ref{sec:char-sub-rates} to the mischaracterized MWPM decoder, which returns the logical class of the most probable fault $\bm f^\star$ under the assumed parameters $\hat{\bm p}$. We assume throughout that the pairs $(p_i, \hat p_i)$ are i.i.d. $\Pi$. 

If the MWPM decoder fails, $\bm f \oplus \bm f^\star$ again contains a simple path $\bm C$ that is a nontrivial logical operator, and the optimality of $\bm f^\star$ now gives
\begin{equation}
    X_{\bm C}(\bm f;\hat{\bm p}) := \frac{\PP[\bm f \oplus \bm C;\hat{\bm p}]}{\PP[\bm f;\hat{\bm p}]} = \prod_{i \in \bm C} \left(\frac{\hat p_i}{1-\hat p_i}\right)^{1-2f_i} \geq 1.
    \label{eq:mis-likelihood}
\end{equation}
The likelihood ratio is evaluated with the assumed $\hat{\bm p}$, but $\bm f$ is drawn with the true $\bm p$. The Chernoff-Cram\'er bound of Eq.~\eqref{eq:chernoff} becomes, for any $s \geq 0$,
\begin{equation}
    \PP_{\bm f}\big[X_{\bm C} \geq 1\big] \leq \prod_{i \in \bm C} B_s(p_i,\hat p_i),
    \label{eq:mis-chernoff}
\end{equation}
with a mischaracterized generalization of the Bhattacharyya coefficient:
\begin{equation}
    B_s(p,\hat p) := (1-p) \left(\frac{\hat p}{1-\hat p}\right)^{s} + p \left(\frac{1-\hat p}{\hat p}\right)^{s}.
    \label{eq:mis-bhat}
\end{equation}
This quantity appears in classical coding theory as a mismatched likelihood ratio moment~\cite{BoroumandM2022} or as a mismatched Bhattacharyya coefficient (when $s=\tfrac12$)~\cite{Alsan2018}.
For $\hat p \neq p$ the $s \leftrightarrow (1-s)$ symmetry of Sec.~\ref{sec:char-sub-rates} is broken, and $s=\tfrac12$ is no longer guaranteed optimal. Averaging over the i.i.d.\ pairs $(p_i,\hat p_i)$ gives $\Ex\left[\PP_{\bm f}(X_{\bm C} \geq 1)\right] \leq \overline B_s[\Pi]^{|\bm C|}$, where $\overline B_s[\Pi] := \Ex_{(p,\hat p) \sim \Pi}[B_s(p,\hat p)]$, and we choose the $s$ that makes this bound tightest,
\begin{equation}
    \overline B[\Pi] := \min_{s \geq 0}\, \overline B_s[\Pi].
    \label{eq:mis-bhat-opt}
\end{equation}
This bounds the asymptotic decay rate, as in Eq.~\eqref{eq:alpha-bracket-kappa}. The path counting arguments for $N(\lambda)$ also carry over unchanged, and in the deep sub-threshold regime,
\begin{equation}
    \alpha[\mathcal D; \Pi] \geq -\ln \overline B[\Pi] - \ln c_1 - c_0\overline B[\Pi] + O\big(\overline B[\Pi]^2\big)
    \label{eq:mis-decay}
\end{equation}
with the same geometric constants $c_0$, $c_1$ and the assumptions we detailed for Eq.~\eqref{eq:char-decay}. Unlike in the characterized case, this informs only the MWPM decoder: a mischaracterized ML decoder may not outperform a mischaracterized MWPM decoder.
This is also only a lower bound: the upper bound remains $-\ln \overline{B}[P]$ as an equivalent expression in $\overline B[\Pi]$ requires the mischaracterized decoder to fail whenever $X_{\bm C}(\bm f;\hat{\bm p}) \geq 1$, which need not hold in general. We nevertheless find numerically that $\alpha$ typically attains this lower bound at small $\overline B[\Pi]$ (see Sec.~\ref{sec:numerics}).

When characterization is perfect, these expressions reduce to those in Sec.~\ref{sec:char-sub}. For a characterized decoder, $\hat p = p$ and $\overline B[\Pi] = \overline B[P]$. For an uncharacterized decoder, $\hat p = \overline p$. Since $B_s$ is linear in $p$, $\overline B_s[\Pi] = B_s(\overline p, \overline p)$, which is minimized at $s=\tfrac12$ to give $B(\overline p)$.

The MWPM decoder depends only on relative weights, so rescaling all weights, $K(\hat p_i) \to cK(\hat p_i)$, cannot change its performance. Since $\big(\hat p/(1-\hat p)\big)^s = e^{-2sK(\hat p)}$, such a rescaling is absorbed by the Chernoff exponent $s \to cs$ (for $c>0$) and leaves $\overline B[\Pi]$ unchanged, as required. A homogeneous $p$ decoded with any homogeneous $\hat p$ gives $\overline B[\Pi] = B(p)$, with the minimum at $s = K(p)/[2K(\hat p)]$.

For an example, consider the log-normal mischaracterization of Sec.~\ref{sec:mis-near}, with $\hat p_i = p_i z_i$, and $p \ll 1$ (neglecting truncation). This gives $B_s(p,\hat p) \approx p^s z^s + p^{1-s} z^{-s}$ and
\begin{equation}
\begin{split}
    \frac{\overline B_{1/2}[\Pi]}{\overline B[P]}
    &\approx \Ex\!\left[\cosh\!\left(\tfrac12\ln z\right)\right] \\ 
    &= e^{\hat\sigma^2/8}\cosh(\hat\sigma^2/4) \\
    &= 1+\frac{\hat\sigma^2}{8}+O(\hat\sigma^4).
\end{split}
\end{equation}
At zero mischaracterization, $\overline B_s[\Pi]$ is minimized at $s=\tfrac12$; optimizing the tilt at nonzero $\hat\sigma$ changes the ratio only at $O(\hat\sigma^4)$. For small $\hat\sigma$, taking the leading logarithmic term of Eq.~\eqref{eq:mis-decay} thus predicts
\begin{equation}
    \alpha[\mathcal D_\mathrm{char};P] - \alpha[\mathcal D_\mathrm{mis};\Pi] \approx \frac{\hat\sigma^2}{8}, \qquad \frac{\Lambda_\mathrm{char}}{\Lambda_\mathrm{mis}} \approx e^{\hat\sigma^2/4}.
    \label{eq:mis-lognormal}
\end{equation}
The effect of mischaracterization here is independent of $\overline p$ (neglecting truncation).
Combining this with the characterization gain of Eq.~\eqref{eq:delta-alpha-lognormal}, for log-normal distributions the mischaracterized decoder outperforms the uncharacterized one by
\begin{equation}
    \alpha[\mathcal D_\mathrm{mis};\Pi] - \alpha[\mathcal D_\mathrm{unchar};P] \approx \frac{\sigma^2 - \hat\sigma^2}{8}.
    \label{eq:mis-vs-unchar}
\end{equation}
To leading order in this model, the predicted gain from characterization is positive when $\hat\sigma^2 < \sigma^2$, matching the near-threshold criterion of Sec.~\ref{sec:mis-near}. For instance, in the example of Sec.~\ref{sec:unchar-decay}, where $\sigma^2 = 0.1$ increases $\Lambda$ by $2.5\%$, estimates with $\hat\sigma^2 = 0.1$ would lose this gain entirely.

Finally, Eq.~\eqref{eq:mis-decay} is an asymptotic result. For a given fault, the output of the MWPM decoder changes discontinuously with $\hat{\bm p}$: it returns the same correction until the weights are perturbed enough to reorder two competing corrections. The smooth Chernoff exponent averages over many such competing paths, and so is accurate only at sufficiently large $d$; we expect deviations at finite distances. As in Sec.~\ref{sec:char-sub}, this section also assumes independent pairs $(p_i,\hat p_i)$. For correlated pairs, $\overline B_s[\Pi]^{|\bm C|}$ is replaced by the joint average of $\prod_{i\in\bm C}B_s(p_i,\hat p_i)$ along each path. Hyperedges and multi-outcome mechanisms disrupt the path-counting arguments and are left as generalizations for future work.

\section{Numerical results}
\label{sec:numerics}

In this section, we report the results of Monte Carlo simulations to investigate misinformed decoding of disordered noise channels on surface codes. 
We find agreement with our theoretical predictions, in particular that across noise and decoding settings, the average entropy approximately sets the threshold of the code, and for deep sub-threshold behavior, the average Bhattacharyya coefficient bounds the decay rate.
For code-capacity noise, ML decoding is performed using the \textsc{planar} decoder of Ref.~\cite{Cao2025}. MWPM decoding uses PyMatching~\cite{Higgott2025sparseblossom}. Importance sampling~\cite{LEcuyer_2009} is used in the deep sub-threshold regime to prioritize noise realizations resulting in logical failures~\cite*{[{Importance sampling algorithm was inspired by the presentation at FTQT: }][]gouzienEstimating2026}. Error bars denote a 95\% confidence interval for logical failure rates across Monte Carlo samples, and two standard errors for fitted parameters; they are omitted from plots when the range is smaller than the marker size. Details for the simulations are provided in Appendix~\ref{app:sup-numerics}.

\subsection{Characterized and uncharacterized decoders}
\label{sec:char-sim}

We first present results on characterized and uncharacterized decoding of surface codes. We find critical average entropies of $H_c = 0.4980(2)$ for code-capacity noise under ML decoding, and $0.19770(6)$ for i.i.d.\ phenomenological, $0.19688(6)$ for time-correlated phenomenological, and $\approx 0.053$ for superconducting-inspired circuit-level noise under MWPM decoding. In the deep sub-threshold regime, MWPM decay rates follow $-\ln\overline{B}[P] -2\overline{B}[P]$ for code-capacity, and $-\ln\overline{B}[P] - 4\overline{B}[P] - 4\overline B[P]^2$ for i.i.d.\ phenomenological noise channels.

\begin{figure}[t]
    \centering
    \vspace{-15pt}
    \includegraphics[width=\linewidth]{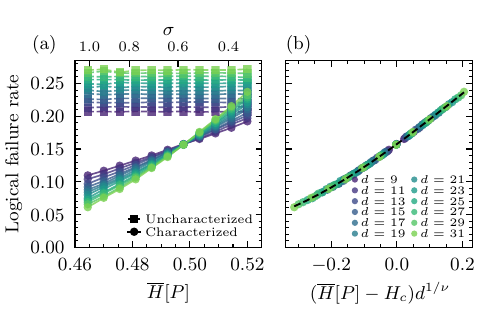}
    \caption{Logical failure rate of a surface code with code-capacity bit-flip noise and ML decoding. (a) Plotted against the average binary entropies $\overline H[P]$ of the error parameter distributions: log-normals with varying $\sigma$ and fixed $\overline{p}=0.12$. Increased disorder lowers the logical failure rate of the characterized decoder.
    (b) Data collapse of the characterized decoder logical failure rate against the scaling variable $(\overline{H}[P]-H_{c})d^{1/\nu}$. See Appendix~\ref{app:sup-numerics} for simulation details.
    }
    \label{fig:LER_vsHbar_bitflip_codecapacity}
\end{figure}

\subsubsection{Code-capacity noise}

Here we simulate a distance $d$ surface code (unrotated planar surface codes unless stated otherwise)~\cite{Bravyi1998}. Each qubit $i$ undergoes the code-capacity bit-flip channel [Eq.~\eqref{eq:noise-channel-ind} with $Q=X_i$] given truncated log-normal error parameters $p_{i}$ [Eq.~\eqref{eq:lognormal}]. Near threshold, we compare the optimal characterized decoder and the optimal uncharacterized decoder, which is the ML decoder initialized with homogeneous error parameters, $p_i = \overline{p}$. We sample from a family of log-normals where the mean is fixed at $\overline{p} = 0.12$, and we sweep $\sigma\in[0.3073,1]$.
The logical failure rates are plotted in Fig.\hyperref[fig:LER_vsHbar_bitflip_codecapacity]{~\ref*{fig:LER_vsHbar_bitflip_codecapacity}(a)} against the parameter distribution's average binary entropy $\overline H[P]$. As $\sigma$ increases, $\overline H[P]$ decreases, as in Eq.~\eqref{eq:delta-H-lognormal}.

The mean $\overline p= 0.12$ is above the optimal threshold for homogeneous noise, $p_{c}\approx 0.1092212(4)$~\cite{Wan2026}.
The uncharacterized decoder is thus above threshold, and its logical failure rates increase with code distance (and remain approximately constant across $\overline{H}[P]$). 
Conversely, for large enough $\sigma$ the average binary entropy, $\overline{H}[P]$, decreases to below the critical value of $H_c \approx \tfrac12$. The characterized decoder can thus operate below threshold, even when $\bar p$ is above the homogeneous threshold. 

We plot a data collapse in Fig.\hyperref[fig:LER_vsHbar_bitflip_codecapacity]{~\ref*{fig:LER_vsHbar_bitflip_codecapacity}(b)} by fitting a third-order polynomial in the scaling variable $(\overline{H}-H_{c})d^{1/\nu}$~\cite{Wang2003}, with data taken from the characterized decoder.
The cubic scaling ansatz returned $H_{c}=0.4980(2)$ and $\nu=1.54(2)$, where uncertainty denotes an approximate $95\%$ statistical confidence interval obtained from the weighted nonlinear least-squares fit, rounded upward to one significant figure. These uncertainties do not include systematic effects from finite-size corrections. Fits restricted to $d=d_{\min},\ldots,31$, with $d_{\min}=9,11,\ldots,25$, showed no clear monotonic drift, and the resulting estimates remained within the quoted intervals. We also tested a quadratic ansatz and a cubic ansatz with an additive correction of $cd^{-\omega}$~\cite{Wang2003}; both gave poorer fits, so we retained the cubic ansatz.
The fitted correlation length critical exponent $\nu$ is consistent with the value obtained for the RBIM at the Nishimori point for both bimodal $\pm J$ disorder and Gaussian-distributed disorder~\cite{Picco2006,Wan2026}.
This agreement suggests that the disordered noise considered here may share the same universality class. We leave a systematic study of which disorder distributions share this critical behavior to future work.

\begin{figure}[t]
    \centering
    \includegraphics[width=\linewidth]{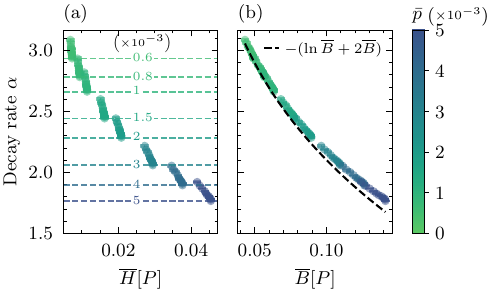}
    \caption{Decay rate $\alpha$ of logical failure rate on $d=13,\ldots,19$ surfaces codes with log-normal code-capacity bit-flip noise and characterized MWPM decoding against (a) average binary entropy (with uncharacterized decoding as dashed lines) and (b) average Bhattacharyya coefficient as we vary the parameter mean $\overline p$ (fixed for each color; annotated) and $\sigma \in [0,1]$. Data collapses onto the predicted $\alpha \approx -\ln\overline{B}[P] -2\overline B[P]$, improving at smaller $\overline B$. See Appendix~\ref{app:sup-numerics} for simulation details.
    }
    \label{fig:alpha-codecapacity-bhat}
\end{figure}

We next consider the deep sub-threshold behavior of these codes by evaluating the decay rate of MWPM decoders, modeling log-normal error parameters at means near $\overline{p}=10^{-3}$.
We simulate families of parameter distributions, defined by holding their $\overline{p}$ fixed while sweeping $\sigma \in [0,1]$. We compute the average logical failure rates for each $P$, before taking the log and estimating $\alpha$ with linear regression. (Using instead the average log failure rate, which captures typical values of $\PP_\mathrm{fail}$ when highly skewed, produces results consistent within $\pm1\%$; see also the discussion of median and mean failure rates in Ref.~\cite{Sriram2025}.) Figure~\hyperref[fig:alpha-codecapacity-bhat]{\ref*{fig:alpha-codecapacity-bhat}(a)} plots these decay rates against the average binary entropy, with uncharacterized decoding indicated by the dashed lines (mean $\overline{p}$ annotated). Figure~\hyperref[fig:alpha-codecapacity-bhat]{\ref*{fig:alpha-codecapacity-bhat}(b)} then plots the same data against the average Bhattacharyya coefficient. We see a saturation of the lower bound of Eq.~\eqref{eq:char-decay}, with $c_0 = 2$ and $c_1 = 1$ for the unrotated surface code with bit-flip noise. At larger $\overline B$, the agreement weakens; this is expected from higher-order powers of $\overline B$ contributing more, and the union bound over-counting via increasingly likely contributions from overlapping paths. Our estimate of $\alpha$ is also an approximation of the true asymptotic $\alpha$ at $d\rightarrow\infty$.

To investigate the applicability of these results to more realistic noise models, we next turn to the phenomenological noise setting, where both physical qubits and stabilizer measurements are considered faulty.

\subsubsection{Phenomenological i.i.d.\ noise}

\begin{figure}[t]
    \centering
    \includegraphics[width=\linewidth]{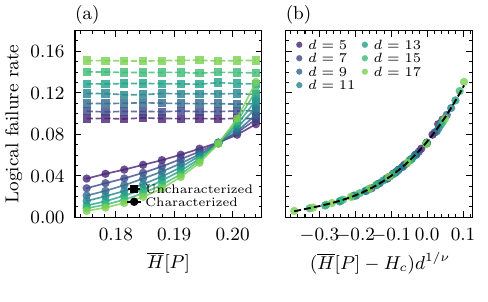}
    \caption{Logical failure rate of a surface code with phenomenological bit-flip and measurement noise using MWPM decoding. (a) Plotted against the average binary entropies $\overline H[P]$ of the error parameter distributions: log-normals with varying $\sigma\in[0.1167,1.1496]$ and fixed $\overline{p}=0.032$. $p_{i}$ are drawn independently at each location in spacetime.
    (b) Data collapse of the characterized decoder logical failure rate against the scaling variable $(\overline{H}[P]-H_{c})d^{1/\nu}$. See Appendix~\ref{app:sup-numerics} for simulation details.
    }
    \label{fig:LER_vsHbar_phenomenological_MWPM}
\end{figure}

In this section, we consider MWPM decoding for both near-threshold and deep sub-threshold simulations, due to its relevance for practical decoding: namely, polynomial time complexity in the system size~\cite{Higgott2025sparseblossom}.
In the code-capacity setting, determining the MWPM is equivalent to computing the ground state of the RBIM~\cite{Amoruso2004}, and can therefore be considered as the zero-temperature limit of the optimal decoding problem.
For the surface code with bit-flip noise, the optimal threshold, i.e. the Nishimori point, is given by $p_{c}\approx 0.1092212(4)$~\cite{Wan2026}. The zero-temperature critical point is given by $p_{c}(T=0)\approx0.103(1)$~\cite{Wang2003,Amoruso2004}. Although the phase boundary is reentrant, these thresholds are comparatively close~\cite{Thomas2011}.
This motivates us to investigate whether the average entropy of the error rate distribution also constrains the threshold of characterized decoders of surface codes with phenomenological noise (which map to a non-self-dual statistical mechanical model) under MWPM decoding (at zero temperature).

We simulate a surface code of distance $d$ with $d$ rounds of (noisy) syndrome extraction. Error parameters for independent $X$ noise channels and measurement errors are drawn independently at each fault location in spacetime from the same parameter distribution $P$. 
For near-threshold simulation, we use a log-normal distribution with $\overline{p}=0.032$, which is above the homogeneous MWPM threshold of $p_{c}\approx0.0293(2)$~\cite{Wang2003}, and sweep $\sigma\in[0.1167,1.1496]$. 
We show in Fig.\hyperref[fig:LER_vsHbar_phenomenological_MWPM]{~\ref*{fig:LER_vsHbar_phenomenological_MWPM}(a)} the logical failure rate versus the average binary entropy.
The uncharacterized decoder is above threshold, and the logical failure increases with code distance but remains constant across $\overline{H}[P]$. For the characterized decoder, we find a crossing in the logical failure rate consistent with a threshold.

We plot a data collapse in Fig.\hyperref[fig:LER_vsHbar_phenomenological_MWPM]{~\ref*{fig:LER_vsHbar_phenomenological_MWPM}(b)} by fitting a third-order polynomial in the scaling variable $(\overline{H}-H_{c})d^{1/\nu}$ to the logical failure rate for the data with $\overline{p}=0.032$. The fit returned values of $H_{c}=0.19770(6)$ and $\nu=1.015(7)$. Again, the fitted correlation length critical exponent $\nu$ is consistent with the value obtained for the zero-temperature critical point of the 3D random plaquette gauge model with homogeneous noise~\cite{Wang2003}.

\begin{figure}[t]
    \centering
    \includegraphics[width=\linewidth]{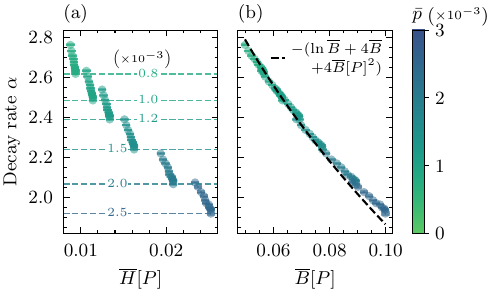}
    \caption{Decay rate $\alpha$ of logical failure rate on $d=15,\ldots,21$ surface codes with i.i.d.\ log-normal phenomenological noise and characterized MWPM decoding against (a) average binary entropy (with uncharacterized decoding as dashed lines) and (b) average Bhattacharyya coefficient as we vary the parameter mean $\overline p$ (fixed for each color; annotated) and $\sigma \in [0,1]$. Data collapses onto the predicted lower bound of $\alpha \gtrsim -\ln\overline{B}[P] -4\overline B[P]-4\overline B[P]^2$. See Appendix~\ref{app:sup-numerics} for simulation details.}
    \label{fig:alpha-phen-bhat}
\end{figure}

We also examine the deep sub-threshold behavior in this phenomenological noise setting. We again plot the decay rate $\alpha$ of surface codes against both the average binary entropy and the average Bhattacharyya coefficient, displayed in Fig.~\ref{fig:alpha-phen-bhat}. We find agreement with the expected form in Eq.~\eqref{eq:char-decay}. For the phenomenological noise model, we have $c_0=4$ instead of $2$, since there are now four steps in the spacetime plane normal to the direction of the logical operator, and the expansion about $\overline B[P]=0$ produces a nonzero quadratic term, $-4\overline B[P]^2$. The collapse onto the predicted line worsens slightly at smaller $\overline{B}[P]$; this is a symptom of the $1/d$ contributions---increasing to larger distances sees the $\alpha$ value approach the line.

\subsubsection{Phenomenological time-correlated noise}

While we observe a good data collapse both near and deep sub-threshold for this i.i.d.\ modeling of $p_i$, this neglects the physically relevant scenario where noise is correlated in time. An example of such time correlations includes two-level systems~\cite{mohseni2026buildquantumsupercomputerscaling}, where a single qubit may have significantly increased physical error rates (but which may still be addressable) that are unchanging over the time scale of a single logical protocol, but vary between experimental shots~\cite{Drouet2026}. To model these correlations, we next simulate a surface code with phenomenological noise where the error parameters for bit-flips and measurement errors $p_{i}$ are drawn at time $t=0$ for each qubit and measurement, and remain fixed for each subsequent time step. The predicted critical entropy is unaffected by these temporal correlations, since only the marginal $P$ enters (see Sec.~\ref{sec:char-near-crit} and Appendix~\ref{app:entropy-corr}), whereas the reduction of the optimal uncharacterized decoder to homogeneous noise no longer holds (see Sec.~\ref{sec:decoders}). We use here a suboptimal uncharacterized decoder: the homogeneous-weight MWPM decoder.

\begin{figure}[t]
    \centering
    \includegraphics[width=\linewidth]{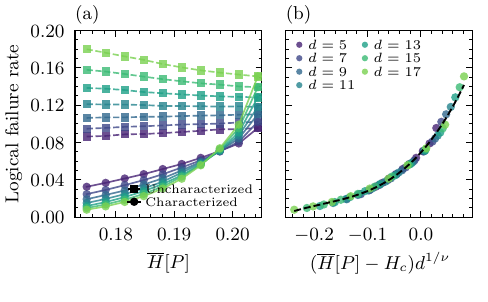}
    \caption{Logical failure rate of characterized and uncharacterized MWPM decoding for surface codes with phenomenological bit-flip and measurement noise, where $p_i$ are drawn at time $t=0$, then set equal across all subsequent $t>0$.
    This time-correlated distribution is log-normal with $\overline{p}=0.032$ and $\sigma$ swept in $[0,1.1496]$. (a) Plotted against the average binary entropy $\overline H[P]$.
    (b) Data collapse of the characterized decoder logical failure rate against the scaling variable $(\overline{H}[P]-H_{c})d^{1/\nu}$. See Appendix~\ref{app:sup-numerics} for simulation details.
    }
    \label{fig:LER_vsHbar_phenomenological_MWPM_fixed_time}
\end{figure}

We show the results in Fig.\hyperref[fig:LER_vsHbar_phenomenological_MWPM_fixed_time]{~\ref*{fig:LER_vsHbar_phenomenological_MWPM_fixed_time}}, where we sweep log-normal distributions with $\sigma\in[0,1.1496]$ for fixed $\overline{p}=0.032$. When $\sigma=0$, the noise is homogeneous, at which point the characterized and uncharacterized logical failure rates coincide (correlations have no effect).
The results are similar to those obtained with $p_{i}$ sampled independently in spacetime, but the uncharacterized decoder, which is above threshold for all distributions, does not have constant logical failure rates as we observed in Fig.\hyperref[fig:LER_vsHbar_phenomenological_MWPM]{~\ref*{fig:LER_vsHbar_phenomenological_MWPM}}. This is expected, as with error parameters fixed in time, the parameter-averaged fault distribution no longer factorizes into a homogeneous channel at $\overline p$.
The data collapse in Fig.\hyperref[fig:LER_vsHbar_phenomenological_MWPM_fixed_time]{~\ref*{fig:LER_vsHbar_phenomenological_MWPM_fixed_time}(b)} is less robust, and deviations from finite-size scaling are visible. The fit to a third-order polynomial in the scaling variable $(\overline{H}-H_{c})d^{1/\nu}$ gave $H_{c}=0.19688(6)$ and $\nu=1.189(9)$. 
The critical entropy differs by approximately $0.4\%$ from the i.i.d.\ value, Fig.~\hyperref[fig:LER_vsHbar_phenomenological_MWPM]{\ref*{fig:LER_vsHbar_phenomenological_MWPM}(b)}. Although numerically close, these estimates have non-overlapping $95\%$ confidence intervals; systematic uncertainty from finite-size corrections is not included. We find that the exponent $\nu$ here is also larger than in the i.i.d.\ case, with non-overlapping confidence intervals.

While the results we derived earlier, including which functionals set the thresholds and decay rates of the codes, appear to hold numerically in the phenomenological noise setting under MWPM decoding, it is a priori unclear to what extent these hold in the more realistic circuit-level noise setting, to which we now turn.

\subsubsection{Circuit-level time-correlated noise}\label{sec:circuit_level_characterized}

\begin{figure*}[t]
    \centering
    \includegraphics[width=\linewidth]{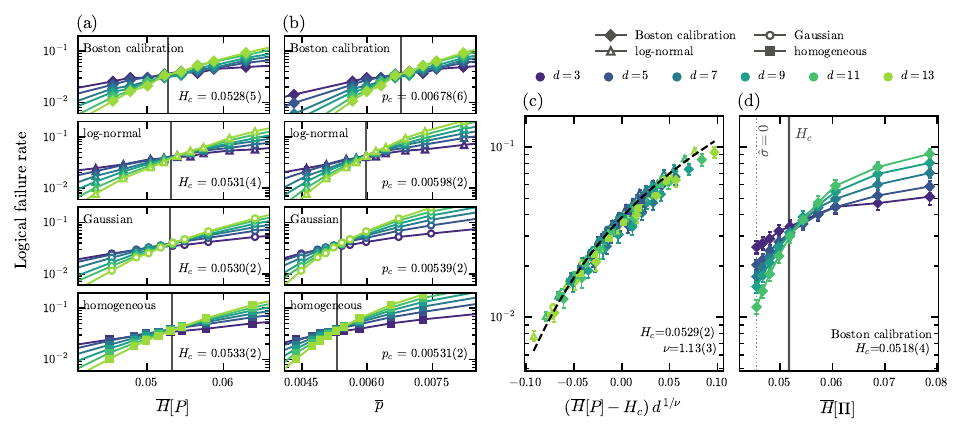}
    \caption{Results from simulated surface code $Z$-memory experiments under circuit-level noise with disordered Pauli noise channels, whose error rates are drawn from one of four distributions: one set by data obtained from the superconducting device \texttt{ibm\_boston} (Boston calibration), along with a log-normal, Gaussian, and homogeneous fit to this device distribution. (a) For each simulation we plot the logical failure rate against the averaged Shannon entropy of the distribution's noise channels, $\overline{H}[P]$ with $H_c$ marked by the gray lines. (b) The same data plotted against the average noise parameter $\overline{p}$ obtained from each distribution. The critical values of $\bar p$ ($p_c$; gray lines) differ much more dramatically. (c) A simultaneous data collapse to a second-order polynomial, for all noise distributions fitting to the same parameters. (d) Results from simulations in which the true noise is set by the Boston calibration data and the decoder assumes mischaracterized noise rates, perturbed with standard deviation $\hat{\sigma}$. The $x$-axis is the average cross-entropy $\overline H[\Pi]$ between the true and decoder distributions.}
    \label{fig:circuit_level_data}
\end{figure*}

In this section, we investigate the circuit-level noise setting, where faults may occur across all elementary operations of an entire specified circuit. We use \emph{Stim}~\cite{Gidney2021} for simulating these noisy circuits. We simulate rotated surface code $Z$-memory experiments with depolarizing circuit-level noise and $d$ rounds of syndrome extraction. We perform decoding with MWPM~\cite{Higgott2025sparseblossom}. We investigate a range of noise distributions, including those that are obtained from calibration data for superconducting devices, available at Ref.~\cite{ibm_compute}. 
We assume perfect temporal correlations, so a fault mechanism retains its drawn error parameter whenever it re-occurs in later rounds.
We observe that the superconducting devices investigated (\texttt{ibm\_boston} and \texttt{ibm\_phoenix}) have noise distributions that differ from log-normality with heavily weighted tails at high-$p$, with results presented in Appendix~\ref{app:CLN_sims}. This corresponds to a small number of components being more noisy than would be expected from a log-normality assumption. This holds both for low-degree and four-valent connectivity devices (\texttt{ibm\_boston} on a heavy-hex lattice, and \texttt{ibm\_phoenix} on a square lattice, respectively). 

We display the results in Fig.~\ref{fig:circuit_level_data}. For the simulations, we draw noise parameters from one of four distributions: a distribution based on the device \texttt{ibm\_boston}, log-normal, Gaussian, and homogeneous. For the device distribution, we construct a Pauli noise model based on the calibration data from \texttt{ibm\_boston} (the distribution from \texttt{ibm\_phoenix} is very similar; see Appendix~\ref{app:CLN_sims}) then obtain a distribution on Pauli noise channels by averaging over six days' worth of calibration data. We then draw from this distribution in our simulations, keeping all noise channels unchanging in time. We then fit log-normal, Gaussian, and homogeneous noise distributions to the \texttt{ibm\_boston} distribution and perform further simulations sampling from these distributions. For the above, we run MWPM as the decoder, using the true (characterized) noise to set the edge weights. We define the average entropy $\overline{H}[P]$ to be the average of the Shannon entropies of all noise channels that appear in a $d=7$ memory circuit run for 7 rounds of syndrome extraction. In this way, the average is weighted by the number of occurrences of each noise channel in spacetime. As can be seen from Fig.~\hyperref[fig:circuit_level_data]{\ref*{fig:circuit_level_data}(a)-(b)}, the fitted values of $H_c$ are in very close agreement between the different noise distributions, unlike what is seen from plotting the logical failure rate against $\overline{p}$. The different values of $H_c$ all lie within $1\%$ of the average, while the critical values of $\bar{p}$ diverge by as much as $15\%$ from the average. Figure~\hyperref[fig:circuit_level_data]{\ref*{fig:circuit_level_data}(c)} shows relatively good, while imperfect, data collapse among all distributions to the same fitted parameters.

Numerical evidence thus reinforces the robustness of the conjecture that the threshold of the code appears to be approximately set by the noise distribution's average entropy, even in this setting (circuit-level noise and MWPM decoding) which differs significantly from the idealized models in which we derived the results. Note that the specific spacetime average of noise channels we perform is only one way of obtaining an average quantity $\overline H [P]$, but may not be the best justified choice. While circuit-level statistical-mechanical mappings are readily constructed~\cite{Pryadko2020, Chubb2021, Aitchison2026}, we have not investigated these models in detail and leave this to future work. We nonetheless still do not expect a consistent collapse onto the Bhattacharyya coefficients in these settings, and we leave its generalization to future work.

\subsection{Mischaracterized decoders}
\label{sec:mis-sim}

A more realistic setting for device noise characterization is the regime where the decoder possesses estimates of the physical error rates, which may or may not be accurate. We now turn to this mischaracterized setting and investigate the effects on the logical performance of decoding surface codes. We present results for bit-flip code-capacity noise and circuit-level noise; we expect the results to hold more generally across other noise settings as in the preceding sections.

\begin{figure}[t]
    \centering
    \vspace{-20pt}
    \includegraphics[width=\linewidth]{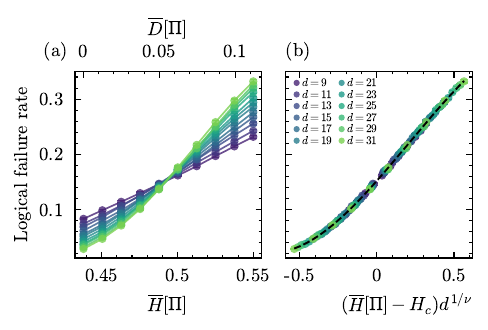}
    \caption{Logical failure rate of a mischaracterized MWPM decoder on a surface code with code-capacity bit-flip noise. (a) Plotted against (top) the average KL divergence $\overline D[\Pi]$ between the true and assumed noise channels, and (bottom) the average binary cross entropy ($\overline{H}[\Pi]=\overline{D}[\Pi]+\overline{H}[P]$). (b) Data collapse of logical failure rate against the scaling variable $(\overline{H}[\Pi]-H_{c})d^{1/\nu}$. See Appendix~\ref{app:sup-numerics} for simulation details.}
    \label{fig:mischar_threshold_code_cap}
\end{figure}

We consider an unrotated surface code with code-capacity bit-flip noise. For each shot, we independently draw each qubit's bit-flip probability $p_{i}$ from a log-normal distribution truncated to $0<p_{i}<1/2$, with post-truncation mean $\overline{p}=0.11$ and width (of $\ln(p)$) given by $\sigma\approx 1.005$. We decode with increasingly mischaracterized MWPM decoders. The results from Sec.~\ref{sec:mis-near} assume ML decoders, and are conjectured to hold only close to the Nishimori line. Nevertheless, we show now that this can remain a useful predictor for MWPM decoders with decoding edge weights sufficiently centered about the true values (at least with surface codes under independent noise, where MWPM decoding is near-optimal). Specifically, after drawing a bit-flip probability $p_{i}$ on each qubit $i$, we decode using MWPM edge weights $2\hat{K}_{i}$ [Eq.~\eqref{eq:nishimori}], which are perturbed additively by a Gaussian,
\[
\hat{K}_{i}=K_{i}+\mathrm{Normal}(0,\hat\sigma^{2}),
\]
with $\hat{\sigma}$ swept from $0$ to approximately $0.645$. 
Using $\hat{p}_{i}=(1+e^{2\hat{K}_{i}})^{-1}$, we then numerically compute
\[
\overline{D}[\Pi]=\frac{1}{N}\sum_{i}\left(p_{i}\log\frac{p_{i}}{\hat{p}_{i}} + (1-p_{i})\log\frac{1-p_{i}}{1-\hat{p}_{i}} \right),
\]
for large Monte Carlo samples $N$.
Observe that $\hat{K}_{i}<0$ corresponds to $\hat{p}_{i}>1/2$.
Varying $\hat{\sigma}$ thus produces joint distributions $\Pi(p,\hat{p})$ with different average KL divergences, while keeping the physical noise distribution fixed.
The results are shown in Fig.~\ref{fig:mischar_threshold_code_cap}. A fit to a third-order polynomial in the scaling variable returned a crossing point of $\overline{H}[\Pi]=0.4926(5)$ bits, and $\nu\approx1.50(3)$. 
This crossing is close to the critical $H_c \approx 0.5$ suggested by the conjecture from Sec.~\ref{sec:mis-near}.

In Fig.~\hyperref[fig:circuit_level_data]{\ref*{fig:circuit_level_data}(d)}, we similarly show that the average cross-entropy approximately sets the threshold in the circuit-level noise setting. We scale the \texttt{ibm\_boston} device noise distribution by 2 (to bring it closer to threshold) and perturb the detector error model given to the MWPM decoder via $p\mapsto z p$ where $z$ is a multiplicative shift with $\ln z \sim \text{Normal}(-\hat\sigma^2/2,\hat\sigma^2)$, which maintains $\mathbb{E}[\hat p|p] = p$. Scaling $\hat\sigma$ leads to a transition at a critical average cross entropy $H_c = 0.0518$, which is close to (within $2.1\%$ of) the critical average entropies seen in the characterized noise distributions.

\begin{figure}[t]
    \centering
    \includegraphics[width=\linewidth]{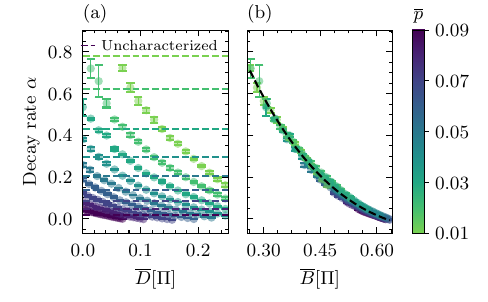}
    \caption{Decay rate $\alpha$ of logical failure rate for mischaracterized MWPM decoders on $d=5,\ldots,13$ surface codes with code-capacity bit-flip noise. (a) Plotted against the average KL divergence $\overline D[\Pi]$ between the true and assumed noise channels. (b) Plotted against the average mischaracterized generalization of the Bhattacharyya coefficient $\overline{B}[\Pi]$, exhibiting data collapse without any fitted parameters. The dashed line is a third-order polynomial fit as a guide to the eye. See Appendix~\ref{app:sup-numerics} for simulation details.}
    \label{fig:alpha_against_KL}
\end{figure}

We next examine mischaracterized decay rates on a surface code with code-capacity bit-flip noise, using log-normal distributions with fixed mutual information $\Delta H[P]\approx 0.018$ (see Appendix~\ref{app:sup-numerics}).
Again, coupling constants are perturbed additively with a Gaussian such that $\hat{K}_{i}=K_{i}+\mathrm{Normal}(0,\hat\sigma^{2})$, where $\hat\sigma\in[0.129,1.88]$.
We then compute the decay rate through linear regression of $-\ln \mathbb{P}_{\mathrm{fail}}$ against $d$, and we plot this in Fig.\hyperref[fig:alpha_against_KL]{~\ref*{fig:alpha_against_KL}(a)}.

The decay rate decreases with increasing divergence, as expected from worsening characterization.
We also overlay the decay rates obtained by an uncharacterized decoder, using homogeneous weights on the decoding graph. Across all distributions, there exists a sufficiently poor mischaracterization with KL divergence $D_{0}$ such that the decay rate obtained by this mischaracterized decoder is smaller than that from an uncharacterized decoder. 
Thus, sufficiently poor mischaracterization of the noise disorder can reduce performance over an uncharacterized decoder.
This is analogous to the below-threshold result in Eq.~\eqref{eq:mis-vs-unchar}.

In Fig.\hyperref[fig:alpha_against_KL]{~\ref*{fig:alpha_against_KL}(b)}, we reparametrize the decay rate by the average mischaracterized generalization of the Bhattacharyya coefficient $\overline{B}[\Pi]$, observing a data collapse without any fitted parameters. We overlay a third-order polynomial in $\overline{B}[\Pi]$ as a guide to the eye. The collapse onto this polynomial in $\overline B$ but not $\overline D$ suggests that the mischaracterized Bhattacharyya coefficient can remain a useful measure of decoding performance even beyond the regime in which it was derived.

\begin{figure*}[t]
    \centering
    \includegraphics{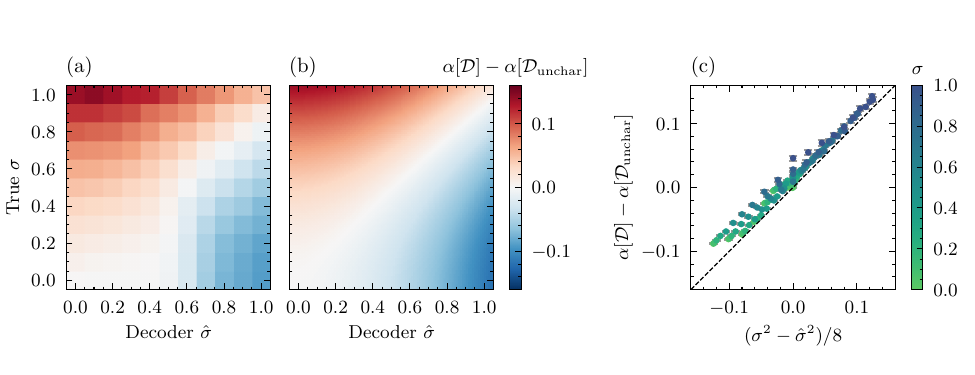}
    \caption{Difference in decay rates of logical failure rates of $d=13,\ldots,19$ surface codes with mischaracterized $\mathcal D$ or uncharacterized $\mathcal D_\mathrm{unchar}$ MWPM decoders and code-capacity bit-flip noise. $p_i$ are drawn from a log-normal distribution with $\overline p=6\times10^{-4}$ and some $\sigma$, while the decoder $\mathcal D$ assumes a perturbed $\hat p_i = p_iz_i$ with $z_i \sim \mathrm{Lognormal}(-\hat\sigma^2/2, \hat\sigma^2)$. At $\sigma=0$ the noise is homogeneous, and at $\hat \sigma = 0$ the noise is perfectly characterized. 
    (a) Numerical estimate from Monte Carlo simulation at $d=13,\ldots,19$. The mischaracterized decoder performs worse than the uncharacterized decoder when $\hat \sigma \gtrsim \sigma$, which is shown in blue. (b) Prediction from Eq.~\eqref{eq:mis-decay} using the mischaracterized generalization of the Bhattacharyya coefficient, Eq.~\eqref{eq:mis-bhat-opt}. (c) The predicted form from Eq.~\eqref{eq:mis-vs-unchar} compared with the simulated data.}
    \label{fig:mischar_grid}
\end{figure*}

We next perform analogous simulations in the deep sub-threshold regime, with i.i.d.\ error parameters drawn from log-normals centered at $\overline p = 6\times10^{-4}$ and with $\sigma\in[0,1]$. The decoder then takes on an assumed $\hat p_i = p_iz_i$, where $z_i$ are i.i.d.\ perturbative log-normals with mean $1$ and $\hat\sigma\in[0,1]$. 
Figure~\ref{fig:mischar_grid} shows the difference in decay rate $\alpha$ between the mischaracterized MWPM decoder $\mathcal D$ and the uncharacterized MWPM decoder $\mathcal D_\mathrm{unchar}$, first for numerical Monte Carlo simulations of MWPM, and then against the predicted expression from Eq.~\eqref{eq:mis-decay} using the mischaracterized Bhattacharyya coefficient. Along the diagonal $\sigma \approx \hat\sigma$, the difference in decay rate starts to disappear, consistent with Eq.~\eqref{eq:mis-vs-unchar}, and exemplified in Fig.\hyperref[fig:mischar_grid]{~\ref*{fig:mischar_grid}(c)}. The numerical data largely agrees with the predicted form, but with a slight positive bias; we anticipate this is due to the effects of distance on $\alpha$, which appear as order $1/d$ corrections. 

\section{Discussion and Outlook}
\label{sec:discussion}

As quantum devices scale to practical demonstrations of quantum error correction, the effects of noise inhomogeneities become increasingly important to understand. Our results establish an information-theoretic framework for understanding decoder performance in the disordered noise setting, with the relevant measure of disorder depending on the operating regime of the code. Near threshold, the average entropy $\overline H[P]$ sets an approximate critical manifold for ML decoders that persists across different families of parameter distributions if correlations are sufficiently small. This includes homogeneous noise, Gaussian and log-normally distributed noise parameters, and even erasure channels. Below threshold, for matchable codes and i.i.d.\ $p_i$, the average Bhattacharyya coefficient $\overline B[P]$ governs bounds on the asymptotic decay rate of MWPM failures, with a simple geometry-dependent form emerging in the deep sub-threshold regime. Consequently, two noise distributions with the same mean or median error rate need not exhibit comparable logical performance; rather, it is approximately set by their $\overline H[P]$ and $\overline B[P]$. These functionals thus provide a concise descriptor of the expected performance of a quantum device, analogous to the simple $p$ in the perfectly characterized homogeneous noise setting.

Disordered noise is not, however, necessarily detrimental to decoding. At fixed mean physical error rate, increasing the spatiotemporal inhomogeneity can improve logical performance, provided the decoder has sufficiently accurate information about the realizations. The average entropy quantifies the additional bits of information available to a decoder per fault location. Uncharacterized decoders that lack access to the realized random error parameters effectively operate at a higher average binary entropy, differing from the characterized decoder precisely by the mutual information. In a mischaracterized setting, the KL divergence between the true and assumed noise channels sets the difference in entropy, and thus predicts the leading-order shift in threshold. Similarly, in the sub-threshold setting, the leading term of the bound on the decay rates [Eq.~\eqref{eq:alpha-bracket-kappa}], $-\ln \overline B[P]$, is related to the Chernoff information of the noise channels. The effects of mischaracterization are summarized by a generalization to the Bhattacharyya coefficient. These results showed that noise disorder, when properly characterized, improves the decay rates and logical performance.

A practical consideration is then the accuracy with which disordered noise must be characterized to be beneficial. Sufficiently large mischaracterization can remove all benefits associated with disordered noise distributions. For log-normal models, our results suggest a particularly simple criterion: characterization is beneficial when the estimation error is smaller than the variation in the underlying true parameter distribution. Otherwise, simply taking an uncharacterized decoder with an accurate mean $\overline p$ is more effective. 

These observations have potential applications to the calibration and benchmarking of quantum devices. Rather than reporting only average or median physical error rates, experimental characterization could additionally report quantities such as the average entropy and Bhattacharyya coefficient of estimated error distributions. For a specified code and decoder, these could provide approximate indicators of proximity to threshold and expected logical error suppression. Moreover, our results also prompt some degree of caution against overly fine-tuned noise characterizations with large levels of uncertainty. For example, for a device with predicted large temporal variations between shots, using sufficiently inaccurate noise estimates for decoder weights can perform worse than using homogeneous weights.

While we have primarily focused on models most suited to quantum devices laid out in a fixed, two-dimensional array, such as superconducting devices, this is not the only setting in which our results are applicable. For example, in photonic setups based on fusion-based quantum computation (FBQC), one has both erasure errors and Pauli errors occurring at different rates~\cite{bartolucci_fusion-based_2023}. The average of the entropies of these noise channels in an FBQC experiment is expected to approximately set the threshold (see Appendix~\ref{app:erasure}). Future work could extend our results to device models that lead to unusual correlation patterns in the noise distributions, such as in trapped ion systems where two-qubit gate error rates are characteristics of ``gate zones'' rather than fixed pairs of qubits~\cite{ransford_98-qubit_2026}, or in neutral atom or spin-qubit devices.

For a given code and noise channel, the result that average entropy is approximately a level set over parameter distributions at criticality opens the door to a method of estimating thresholds that may be otherwise difficult to obtain numerically: the erasure threshold is set by the percolation transition at the critical erasure rate $q_c$ on the dual lattice, which is often readily computable. The critical condition then predicts that all other parameter distributions should attain threshold near $\overline H[P_c] \approx q_c$. Future work could examine further the applicability of this method, and the level of accuracy it can attain in predicting ML thresholds across different codes, circuits, and noise channels.

Several limitations remain. Although our threshold results cover error parameters with mild spatial and extended temporal correlations (such that macroscopic spatial defects are suppressed), the sub-threshold decay rate results assume each $p_i$ is i.i.d., including in the phenomenological noise setting. Natural extensions would therefore include extending the sub-threshold decay rate results to settings with correlations, and to also consider processes with classical memory (\textsf{CM}) outside of the $\textsf{MM}$ setting, such that the environment records information about the system and feeds it forward to condition later noise. This would model, for example, error rates that increase following earlier errors due to heating, leakage, or cosmic ray events. Our results do not straightforwardly apply to non-Pauli noise processes which may better describe the noise encountered in real devices~\cite{Ware_2021,Hashim_2021,Ficheux_2018}. In particular, our claims of the benefits of spatial noise inhomogeneity should be tempered by the reality that noisy components in real devices may result in larger leakage rates which can be highly detrimental to error-correction experiments~\cite{Brown_2019,Miao_2023,He_2025}.   

In addition, the principal Boltzmann factor prediction for thresholds is conjectural and uncontrolled, even in the case of homogeneous noise. A rigorous proof that the critical points predicted by the cluster construction of Ref.~\cite{Ohzeki2009} converge to the true transition point as the cluster size tends to infinity would be desirable. It would be interesting to see if a simple functional emerges, which may involve higher-order moments of the error parameter distributions, when taking such an approach.

Future work could integrate the functionals identified in this work into decoders which calibrate during operation, where these quantities could be reported as useful soft information to the decoder.
The results in this work could be applied to more general noise channels through extensions to the statistical mechanical models to which coherent noise~\cite{vennCoherentError2023, ecksteinRobust2024, ecksteinLearning2025, behrendsSurface2025, Behrends2025, negariLongrange2026, yanNonlinear2026, yangDecoding2026, bejanProjected2026}, and correlated noise~\cite{Chubb2021} map. It would also be of interest to generalize the results presented in this work to the problem of sliding window decoding~\cite{Dennis2002, huangShor2021, skoricParallel2023, tanScalable2023, Sriram_2026}, and in particular, in the presence of physical error parameter distributions that drift in time~\cite{Bhardwaj2026,Proctor2020,Pelofske2023,Wang2024,Kim2025,Gaye2025}.

While we focused on surface codes in various noise settings in this work, the results should extend to more general codes---including quantum low-density parity check codes---which map to disordered statistical mechanical models~\cite{Kovalev2015, Chubb2021}.
Additionally, while MWPM and ML decoding correspond to well-defined operations on the mapped statistical mechanical model~\cite{Chubb2021}, we do not expect generic heuristic decoders, without such a thermodynamic interpretation, to satisfy the results presented in this work. We leave an investigation of such decoders, including those which utilize belief propagation~\cite{Gallager1963,Panteleev2021}, to future work.

\begin{acknowledgments}
We thank Timo Hillmann and Stephen Bartlett for helpful discussions. 
This research was prompted by conversations at the FTQT 2026 conference in Benasque.
CTA is supported by the Gates Cambridge Trust. BB acknowledges funding from
EPSRC grant EP/V062654/1. 
DJW is supported by the Australian Research Council Discovery Early Career Research Award (DE220100625). CKM acknowledges support from the Intelligence
Advanced Research Projects Activity (IARPA), under
the Entangled Logical Qubits program through Cooperative Agreement Number W911NF-23-2-0223. This research was conducted using resources of the National Computational Infrastructure HPC `Gadi', an NCRIS-enabled capability supported by the Australian Government. Access to these resources was provided by the Sydney Informatics Hub, a Core Research Facility of The University of Sydney, as supported by the Deputy Vice-Chancellor (Research). This work was also performed using resources provided by the Cambridge Service for Data Driven Discovery (CSD3) operated by the University of Cambridge Research Computing Service, provided by Dell EMC and Intel using Tier-2 funding from the Engineering and Physical Sciences Research Council (capital grant EP/T022159/1), and DiRAC funding from the Science and Technology Facilities Council. We also acknowledge the use of the `Swirles' High Performance Computing cluster at the Faculty of Mathematics, University of Cambridge.
\end{acknowledgments}

\bibliography{references.bib}

\onecolumngrid
\appendix

\section{Ordering of disordered noise channels}
\label{app:coding-theory}

In this section we use tools from classical coding theory to order disordered noise channels by their decoding performance, focusing on characterized ML decoding for i.i.d.\ $p_i$ with binary noise channels (unless specified otherwise). We show that disordered noise channels tend to result in better decoder performance: at fixed $\bar p$, increasing the disorder increases the thresholds of characterized decoders; meanwhile, at fixed $\overline H [P]$, disordered noise results in greater sub-threshold error suppression. We shall use the notation
\begin{equation}
    \overline\PP_\mathrm{fail}^\mathrm{ML}[P] := \Ex_{\bm p\sim P}\Big[\PP_\mathrm{fail}\big[\mathcal D_\mathrm{ML}^{\bm p};\bm p\big]\Big]
\end{equation}
for the average logical failure rate of the characterized ML decoder under the parameter distribution $P$, for a fixed code or circuit and noise channels.

\subsection{Convex ordering and degradation}

For two parameter distributions $P$ and $P'$, we say that $P$ is smaller than $P'$ in the convex order, $P \cx P'$, if
\begin{equation}
    \Ex_{p\sim P}[g(p)] \leq \Ex_{p\sim P'}[g(p)] \quad \text{for all convex functions } g,
\end{equation}
with the inequality reversed for all concave $g$. Taking $g(p) = \pm p$ shows that $P$ and $P'$ have equal means; $P'$ is, however, more ``spread out'' than $P$. Since the binary entropy $H(p)$ and Bhattacharyya coefficient $B(p)$ are both concave in $p$ in the domain $p \in (0,\tfrac12)$, we also have that $\overline H[P] \geq \overline H[P']$ and $\overline B[P] \geq \overline B[P']$.

A related notion is that of channel degradation. We regard a disordered noise channel as outputting, at each location, both the fault $f_i$ and its error parameter $p_i$. (The faults would then be mapped onto some syndrome, which is passed to the decoders.) A channel $W$ is degraded with respect to a channel $V$, written $W \degr V$, if the output of $W$ can be simulated by applying further randomization to the output of $V$, independently of the input. Degradation implies that the optimal logical failure rate under $W$ is at least that under $V$~\cite{Richardson2008}.

Notably, convex ordering implies degradation (cf.\ Ref.~\cite{Alsan2013}, where a related convex ordering of binary channels is shown to be equivalent to degradation for symmetric channels). Let $V$ and $V'$ be disordered noise channels with parameter distributions $P \cx P'$. By Strassen's theorem~\cite{Strassen1965}, there is a joint distribution $\pi(p,p')$ with marginals $P$ and $P'$ such that $\Ex_\pi[p'\,|\,p] = p$. Given an output $(\bm f,\bm p')$ of $V'$, we draw each $p_i$ from the conditional distribution $\pi(p_i\,|\,p_i')$ and discard $\bm p'$. By construction, $\bm p \sim P$. Moreover, $f_i$ depends on $p_i$ only through $p_i'$, so conditioned on $p_i$ it is Bernoulli with parameter $\Ex_\pi[p_i'\,|\,p_i] = p_i$, independently across locations $i$. The resulting pair $(\bm f,\bm p)$ is therefore distributed exactly as the output of $V$, so $V \degr V'$. Since this simulation leaves the fault, and hence its syndrome and logical class, unchanged, any decoder for $V$ yields a (randomized) decoder for $V'$ with the same failure rate. Hence, for any code or circuit,
\begin{equation}
    P \cx P' \quad\Longrightarrow\quad \overline\PP_\mathrm{fail}^\mathrm{ML}[P] \geq \overline\PP_\mathrm{fail}^\mathrm{ML}[P'].
    \label{eq:degradation-order}
\end{equation}
That is, at fixed mean, spreading the parameter distribution can only decrease the average logical failure rate of the characterized ML decoder.

\subsection{Ordering at fixed mean}

At fixed mean $\overline p$, the homogeneous and erasure distributions are the extremes of the convex order. To see this, note that for any $P$ with mean $\overline p$ and any convex $g$ on $[0,\frac12]$, Jensen's inequality gives $g(\overline p) \leq \Ex_{p\sim P}[g(p)]$. Conversely, a convex function lies below its chord between $p=0$ and $p=\frac12$, so that $g(p) \leq (1-2p)\,g(0) + 2p\, g(\tfrac12)$ on $[0,\frac12]$; taking the expectation gives $\Ex_{p\sim P}[g(p)] \leq (1-2\overline p)\,g(0) + 2\overline p\, g(\tfrac12)$. Together, we get
\begin{equation}
    \delta(p-\overline p) \ \cx\ P \ \cx\ P_{\mathrm{er},2\overline p}, \qquad P_{\mathrm{er},2\overline p}(p) = (1-2\overline p)\,\delta(p) + 2\overline p\,\delta\big(p-\tfrac12\big),
\end{equation}
with $P_{\mathrm{er},q}$ the erasure distribution of Sec.~\ref{sec:char-near}. If $P$ and $P'$ are both log-normal with equal mean but $\Var[P] \leq \Var[P']$, then we also have $P \cx P'$. For example, with $P\sim\mathrm{lognormal}(\mu-\sigma^2/2,\sigma^2)$, increasing $\sigma^2$ results in smaller $\overline H[P]$ and $\overline B[P]$.

This has consequences for the logical failure rates: by Eq.~\eqref{eq:degradation-order}, for any code or circuit,
\begin{equation}
    \overline\PP_\mathrm{fail}^\mathrm{ML}\big[P_{\mathrm{er},2\overline p}\big] \leq \overline\PP_\mathrm{fail}^\mathrm{ML}[P] \leq \overline\PP_\mathrm{fail}^\mathrm{ML}\big[\delta(p-\overline p)\big].
    \label{eq:mean-order}
\end{equation}
Let $p_c$ be the threshold under homogeneous noise and $q_c$ the erasure threshold. The homogeneous channel at $\overline p$ is below threshold for $\overline p < p_c$, and the erasure channel $P_{\mathrm{er},2\overline p}$ is above threshold for $2\overline p > q_c$. Equation~\eqref{eq:mean-order} then bounds every $P$: the characterized ML decoder is below threshold whenever $\overline p < p_c$, and above threshold whenever $\overline p > q_c/2$.

For self-dual models such as the toric code under bit-flip noise, $p_c \approx 0.109$ and $q_c = \tfrac12$~\cite{English2026}. The threshold of any disordered noise channel, in terms of its mean error parameter, therefore lies between $0.109$ and $0.25$. Somewhat counter-intuitively, the homogeneous threshold of $0.109$ is the \emph{worst} possible threshold in the mean over all parameter distributions. Heuristically, this is because every other distribution has some probability of realizing error parameters below $\overline p$; a characterized decoder can exploit these reliable locations, pushing down the failure rate even as the mean error parameter increases. An uncharacterized decoder cannot, and its threshold remains at $\overline p = p_c$ for every $P$ (Appendix~\ref{app:bayes}). Equation~\eqref{eq:mean-order} is the exact counterpart of the Jensen inequality $\overline H[P] \leq H(\overline p)$ of Sec.~\ref{sec:unchar-threshold}: both favor spread-out distributions over homogeneous noise at the same mean, but Eq.~\eqref{eq:mean-order} holds at every code distance, without the replica assumptions of Sec.~\ref{sec:char-near}.

\subsection{Ordering at fixed average entropy}

Instead of the distribution of $p$, consider the distribution $P_H$ of the binary entropy $h = H(p)$ for $p\sim P$, which has mean $\overline H[P]$. Convex ordering in $P_H$ then orders the deep sub-threshold behavior at fixed average entropy. The average Bhattacharyya coefficient can be written as $\overline B[P] = \Ex_{h\sim P_H}[\beta(h)]$, with $\beta(h) := B\big(H^{-1}(h)\big)$ and $H^{-1}$ the inverse of $H$ on $[0,\frac12]$. The function $\beta$ is strictly concave on $[0,1]$, the support of $h$. To see this, use the natural logarithm in $H$ (which rescales $h$ and does not affect concavity), and write $u = p(1-p)$ and primes for derivatives with respect to $p$:
\begin{equation}
    \frac{\diff^2\beta}{\diff h^2} = \frac{B''H' - B'H''}{(H')^3} = \frac{1}{u^{3/2}(H')^3}\left[(1-2p) - \frac12\ln\frac{1-p}{p}\right].
\end{equation}
Here $H' = \ln[(1-p)/p] > 0$ on $(0,\frac12)$, and, with $t = 1-2p \in (0,1)$, $\frac12\ln[(1-p)/p] = \operatorname{arctanh} t > t$, so the second derivative is negative. For two parameter distributions $P$ and $P'$ with entropy distributions $P_H \cx P'_H$ (so that $\overline H[P] = \overline H[P']$), concavity of $\beta$ then gives
\begin{equation}
    \overline B[P] \geq \overline B[P'].
    \label{eq:entropy-order}
\end{equation}
At fixed average binary entropy, the average Bhattacharyya coefficient is therefore largest for homogeneous noise, $P_H(h) = \delta(h-\overline H)$. It decreases under convex-order spreading of the entropy distribution, and is smallest for erasures, for which $P_H$ has mass only at $h = 0$ and $h = 1$ and $\overline B = \overline H$. As per Eq.~\eqref{eq:char-decay}, a smaller $\overline B[P]$ increases the lower bound on the decay rate, suggesting an improvement to decoding performance wherever that bound applies.

\section{Optimal uncharacterized decoders}
\label{app:bayes}

In Sec.~\ref{sec:decoders} we reasoned that the optimal uncharacterized decoder is the ML decoder for the homogeneous channel at the mean error parameter $\overline p$. Here we derive this result, and show that it depends on how decoding performance is compared across realizations of $\bm p$: other natural figures of merit lead to different optimal decoders.

An uncharacterized decoder $\mathcal D$ takes only the syndrome as input; it may depend on the parameter distribution $P$, but not on the realized $\bm p$. We compare it with the characterized ML decoder $\mathcal D_\mathrm{ML}^{\bm p}$, whose failure rate we abbreviate as $\PP_\mathrm{fail}^\mathrm{ML}[\bm p] := \PP_\mathrm{fail}[\mathcal D_\mathrm{ML}^{\bm p};\bm p]$. We also write
\begin{equation}
    \PP[\bm{\mathsf s},\ell;\bm p] := \sum_{\bm f\,:\,\mathsf s(\bm f)=\bm{\mathsf s},\,\ell(\bm f)=\ell} \PP[\bm f;\bm p]
\end{equation}
for the probability of the logical class $\ell$ with syndrome $\bm{\mathsf s}$ [proportional to $\mathcal Z_{\ell,\bm{\mathsf s}}[\bm p]$, Eq.~\eqref{eq:partition}], and define the Bayes decoder for a (not necessarily factorized) prior distribution $Q(\bm p)$ as
\begin{equation}
    \mathcal D_\mathrm{Bayes}^{Q}[\bm{\mathsf s}] := \argmax_\ell\, \Ex_{\bm p\sim Q}\big[\PP[\bm{\mathsf s},\ell;\bm p]\big].
    \label{eq:bayes-decoder}
\end{equation}

\subsection{Average failure rate}

The simplest figure of merit is the average excess failure rate relative to the characterized ML decoder,
\begin{equation}
    M_\mathrm{diff}[\mathcal D] := \Ex_{\bm p\sim P}\Big[\PP_\mathrm{fail}[\mathcal D;\bm p] - \PP_\mathrm{fail}^\mathrm{ML}[\bm p]\Big] \geq 0.
\end{equation}
The second term does not depend on $\mathcal D$, so minimizing $M_\mathrm{diff}$ is equivalent to minimizing the average failure rate $\overline\PP_\mathrm{fail}[\mathcal D;P]$ of Sec.~\ref{sec:decoders}. Since $\mathcal D$ does not depend on $\bm p$, the average can be moved inside the sum over faults:
\begin{align}
    \overline\PP_\mathrm{fail}[\mathcal D;P] &= \Ex_{\bm p\sim P}\bigg[1 - \sum_{\bm f} \PP[\bm f;\bm p] \ \delta_{\mathcal D[\textsf s(\bm f)] - \ell(\bm f)}\bigg] \nonumber\\
    &= \sum_{\bm f} \Ex_{\bm p\sim P}\big[\PP[\bm f;\bm p]\big]\ \left[1 - \delta_{\mathcal D[\textsf s(\bm f)] - \ell(\bm f)}\right].
\end{align}
This is the failure rate under the marginal fault distribution, which for i.i.d.\ $p_i$ is the homogeneous distribution at $\overline p$,
\begin{equation}
    \Ex_{\bm p \sim P} \big[\PP[\bm f;\bm p] \big] = \prod_{i=1}^n \overline p ^{f_i}(1-\overline p)^{1-f_i}.
\end{equation}

The optimal uncharacterized decoder for i.i.d.\ $p_i$ is then the Bayes decoder for the prior $P$, which coincides with the ML decoder for homogeneous noise at $\overline p$, $\mathcal D_\mathrm{Bayes}^{P} = \mathcal D_\mathrm{ML}^{\overline p}$. For any uncharacterized decoder $\mathcal D$,
\begin{equation}
    \overline\PP_\mathrm{fail}[\mathcal D;P] \geq \overline\PP_\mathrm{fail}\big[\mathcal D_\mathrm{ML}^{\overline p};P\big] = \PP_\mathrm{fail}\big[\mathcal D_\mathrm{ML}^{\overline p};\overline p\big],
\end{equation}
where the right-hand side is the failure rate of the ML decoder under homogeneous noise at $\overline p$. No uncharacterized decoder can outperform, on average, the optimal decoder for homogeneous noise with the same mean; in particular, the two share the same threshold as a function of $\overline p$. The same decoder also minimizes the ratio of averages, $\overline\PP_\mathrm{fail}[\mathcal D;P] / \Ex_{\bm p\sim P}\big[\PP^\mathrm{ML}_\mathrm{fail}[\bm p]\big]$, since the denominator does not depend on $\mathcal D$.

Otherwise, with $p_i$ drawn once for each physical fault mechanism $i=1,\ldots,n'$ and then fixed in time $t=1,\ldots,T$, we get an expression of the form
\begin{align}
    \Ex_{\bm p \sim \mathcal P} \big[\PP[\bm f;\bm p] \big] =\prod_{i=1}^{n'} \Ex_{p\sim P}\left[p ^{\sum_{t=1}^T f_{it}}(1-p)^{\sum_{t=1}^T (1-f_{it})}\right],
\end{align}
which is not expressible as the mean of the distribution. The optimal uncharacterized decoder is then the Bayes decoder $\mathcal D_\mathrm{Bayes}^{\mathcal P}$ for this non-factorized prior, which can exploit the correlations (for example, by inferring persistently large error parameters from the syndrome history) and so outperform $\mathcal D_\mathrm{ML}^{\overline p}$. 

\subsection{Average ratio}

The average failure rate is dominated by realizations of $\bm p$ with large failure rates. If we instead weight all realizations equally on a relative scale, using the average ratio (assuming nonzero failure probabilities)
\begin{equation}
    M_\mathrm{ratio}[\mathcal D] := \Ex_{\bm p\sim P}\left[\frac{\PP_\mathrm{fail}[\mathcal D;\bm p]}{\PP_\mathrm{fail}^\mathrm{ML}[\bm p]}\right] \geq 1,
\end{equation}
then $\mathcal D_\mathrm{ML}^{\overline p}$ is no longer optimal. Defining the tilted prior
\begin{equation}
    \widetilde P(\bm p) := \frac{1}{\widetilde{\mathcal Z}}\, \frac{P(\bm p)}{\PP_\mathrm{fail}^\mathrm{ML}[\bm p]}, \qquad \widetilde{\mathcal Z} := \Ex_{\bm p\sim P}\left[\frac{1}{\PP_\mathrm{fail}^\mathrm{ML}[\bm p]}\right],
\end{equation}
and assuming $\widetilde{\mathcal Z}$ is finite, we have $M_\mathrm{ratio}[\mathcal D] = \widetilde{\mathcal Z}\;\overline\PP_\mathrm{fail}[\mathcal D;\widetilde P]$. By the argument above, this is minimized by the Bayes decoder for the tilted prior, $\mathcal D_\mathrm{Bayes}^{\widetilde P}$. The tilt favors realizations of $\bm p$ on which ML decoding succeeds, i.e.\ those with smaller error parameters. Since $\widetilde P$ does not factorize over locations, $\mathcal D_\mathrm{Bayes}^{\widetilde P}$ is not in general an ML decoder for any homogeneous channel.

The size of this effect can be estimated below threshold. There, $\PP_\mathrm{fail}^\mathrm{ML}[\bm p]$ is dominated by short logical paths, $\PP_\mathrm{fail}^\mathrm{ML}[\bm p] \sim \sum_{\bm C}\prod_{i\in\bm C} B(p_i)$ (Sec.~\ref{sec:char-sub-rates}). For $p_i \ll 1$, $B(p_i) \approx 2\sqrt{p_i}$, so $\partial \ln \PP_\mathrm{fail}^\mathrm{ML}/\partial \ln p_i \approx w_i/2$, where $w_i$ is the fraction of the failure probability carried by paths through location $i$. Since $\sum_i w_i$ is the typical length of a failing path, of order $d$, a typical location has $w_i \approx d/n$. To leading order, the tilt reweights the distribution of each $p_i$ by $p_i^{-w_i/2}$; for log-normal $P$ this lowers its mean by a relative amount $\approx w_i\sigma^2/2 \approx O(d\sigma^2/n)$. This suggests a small per-location shift for $n \gg d$. The direction of the shift is, moreover, a consequence of the choice of figure of merit: the analogous ratio of success rates, $1-\PP_\mathrm{fail}$, tilts the prior towards larger error parameters instead.

\subsection{Geometric mean}

Below threshold, the failure rate at a fixed realization scales as $\PP_\mathrm{fail}[\mathcal D;\bm p] \approx e^{-\alpha_{\mathcal D}(\bm p)\,d}$, which suggests comparing decoders through the geometric mean
\begin{equation}
    M_\mathrm{geom}[\mathcal D] := \exp \left\{\Ex_{\bm p\sim P}\left[\ln \frac{\PP_\mathrm{fail}[\mathcal D;\bm p]}{\PP^\mathrm{ML}_\mathrm{fail}[\bm p]} \right]\right\} \approx e^{(\overline\alpha_\mathrm{ML} - \overline\alpha_{\mathcal D})\,d} \geq 1,
\end{equation}
where $\overline\alpha_{\mathcal D} := \Ex_{\bm p\sim P}[\alpha_{\mathcal D}(\bm p)]$. Since $\Ex[\ln X] \leq \ln \Ex[X]$, a geometric mean is less sensitive than an arithmetic mean to rare realizations of $\bm p$ with large failure rates; when $\ln \PP_\mathrm{fail}$ concentrates about its mean, as expected for large codes, it tracks the typical rather than the average failure rate (the analogue of a quenched rather than annealed average). The optimal decoder satisfies
\begin{equation}
    \mathcal D_\mathrm{geom} = \argmin_{\mathcal D}\, \Ex_{\bm p\sim P} \left[\ln\PP_\mathrm{fail}[\mathcal D;\bm p]\right].
    \label{eq:D_geom}
\end{equation}
The average failure rate is a sum over syndromes of terms that each depend only on the decoder's output for that syndrome, so it can be minimized one syndrome at a time. Taking the logarithm couples the syndromes: the best output for one syndrome now depends on the outputs for all the others. We can nevertheless derive a condition that any optimal decoder must satisfy. Let $\widehat{\mathcal D}$ agree with $\mathcal D_\mathrm{geom}$ on every syndrome except one, $\bm{\mathsf s}$, where it returns $\hat\ell$ instead of $\ell = \mathcal D_\mathrm{geom}[\bm{\mathsf s}]$. The failure rate changes by
\begin{equation}
    \gamma(\bm p) := \PP_\mathrm{fail}[\widehat{\mathcal D};\bm p] - \PP_\mathrm{fail}[\mathcal D_\mathrm{geom};\bm p] = \PP[\bm{\mathsf s},\ell;\bm p] - \PP[\bm{\mathsf s},\hat\ell;\bm p].
\end{equation}
Optimality of $\mathcal D_\mathrm{geom}$ and the concavity of the logarithm, $\ln(1+x) \leq x$, then give
\begin{equation}
    0 \leq \Ex_{\bm p\sim P}\left[\ln\left(1 + \frac{\gamma(\bm p)}{\PP_\mathrm{fail}[\mathcal D_\mathrm{geom};\bm p]} \right) \right] \leq \Ex_{\bm p\sim P} \left[\frac{\PP[\bm{\mathsf s},\ell;\bm p]-\PP[\bm{\mathsf s},\hat\ell;\bm p]}{\PP_\mathrm{fail}[\mathcal D_\mathrm{geom};\bm p]} \right]
\end{equation}
for every $\hat\ell$. Since this holds for every syndrome, $\mathcal D_\mathrm{geom}$ must satisfy the self-consistency condition
\begin{equation}
    \mathcal D_\mathrm{geom} = \mathcal D_\mathrm{Bayes}^{P^\star}, \qquad P^\star(\bm p) := \frac1{\mathcal Z^\star}\, \frac{P(\bm p)}{\PP_\mathrm{fail}[\mathcal D_\mathrm{geom};\bm p]},
\end{equation}
with $\mathcal Z^\star$ a normalization constant. That is, $\mathcal D_\mathrm{geom}$ is the Bayes decoder for a prior tilted by its own failure rate, which suggests solving for it iteratively. The converse need not hold: a decoder satisfying this condition is not guaranteed to minimize $M_\mathrm{geom}$, and the condition may have several solutions. As for the average ratio, the tilt favors realizations of $\bm p$ on which the decoder succeeds, i.e.\ those with small error parameters. Heuristically, $\mathcal D_\mathrm{geom}$ therefore behaves like a decoder that assumes lower error parameters than $\mathcal D_\mathrm{ML}^{\overline p}$, and so lies closer to the MWPM decoder.

\section{Entropic analysis}
\label{app:entropy}

Here we explain in more detail the entropic level-set conjecture from Sec.~\ref{sec:char-near}, which expands the results from Refs.~\cite{Nishimori2002, Takeda2005} to the context of disordered decoding of QEC codes on the Nishimori line. We first derive the simple case using binary faults and i.i.d.\ $p_i$. We consider correlated error parameters in Appendix~\ref{app:entropy-corr}, mechanisms with more than two outcomes (such as depolarizing noise) in Appendix~\ref{app:entropy-corr-noise}, and off-Nishimori generalizations in Appendix~\ref{app:entropy-off-nish}.

We start with a normalized version of the partition function from Eq.~\eqref{eq:partition}, which avoids singularities when $p=0$ has nonzero mass: 
\begin{equation}
    \widetilde{\mathcal Z}[\bm f;\bm p] := \left(\prod_{i=1}^n \sqrt{p_i(1-p_i)}\right) \mathcal Z[\bm f;\bm p].
\end{equation}
Since the prefactor depends on $\bm p$ but not $\bm f$, this preserves the free energy differences of logical classes.
We consider $N$ replicas of this partition function, with shared (quenched) reference fault $\bm f$ and shared error parameters $\bm p$. For i.i.d.\ $p_i$, the disorder average over $\bm f$ and $\bm p$ factorizes over mechanisms, giving
\begin{align}
    \Ex_{\bm p \sim P}\left[\left\langle \widetilde{\mathcal Z}[\bm f;\bm p]^N \right\rangle_{\bm f}\right]
    &= \int\!dp_1 \cdots\int\!dp_n \prod_{i=1}^n P(p_i)[p_i(1-p_i)]^{N/2}\ \sum_{\bm f} \PP[\bm f;\bm p] \prod_{r=1}^N \sum_{\bm S^{(r)}} e^{-\mathcal H[\bm S^{(r)}\!\!, \bm f;\bm p]}\\
    &= \sum_{\{\bm S^{(r)}\}_{r=1}^N} \prod_{i=1}^n \int dp_i P(p_i)[p_i(1-p_i)]^{N/2} \sum_{f_i=0}^1 \PP[f_i;p_i]\prod_{r=1}^N e^{K(p_i)\eta(f_i) \prod_k \left(S_k^{(r)}\right)^{(\bm g_k)_i}} \\
    &= \sum_{\{\bm S^{(r)}\}_{r=1}^N} \prod_{i=1}^n x(w_i;P) \\
    &:= \overline{\mathcal Z}^N[x(0;P), x(1;P),\ldots, x(N;P)]
\end{align}
with averaged Boltzmann factors
\begin{equation}
    x(w;P) = \Ex_{p\sim P}\left[(1-p)^{N-w+1}p^w + p^{N-w+1}(1-p)^w\right],
    \label{eq:Boltz-x}
\end{equation}
where $w_i$ is the number of replicas with odd parity spin product at interaction $i$:
\begin{equation}
    w_i = \sum_{r=1}^N \frac12 \left[ 1- \prod_k \left(S^{(r)}_k\right)^{(\bm g_k)_i}\right].
\end{equation}
By the same construction applied to the Kramers-Wannier dual model, with dual replica spins $\bm S^{*(r)}$ and dual weights $w_i^*$ defined analogously, the dual model's replicated partition function is
\begin{equation}
    \overline{\mathcal Z}^{*N}\!\left[x^*(0;P),\ldots,x^*(N;P)\right] := \sum_{\{\bm S^{*(r)}\}_{r=1}^N} \prod_{i=1}^n x^*(w_i^*;P),
\end{equation}
with $\mathbb Z_2^N$ Fourier transformed Boltzmann factors
\begin{equation}
    x^*(w^*;P) = \Ex_{p\sim P}\bigg[ 2^{-N/2} (1-2p)^{w^*}\left([1-p] + (-1)^{w^*}p\right)\bigg] 
    \label{eq:Boltz-x-dual}
\end{equation}
for $0 \leq w_i^* \leq N$. By substituting in the above expressions, and letting $P^*$ be some disorder distribution of the model on the dual lattice, the averaged partition function and the averaged partition function of the dual model satisfy~\cite{Takeda2005} 
\begin{equation}
    \overline{\mathcal Z}^N\!\left[x(0;P),\ldots,x(N;P)\right] \ \overline{\mathcal Z}^{*N}\!\left[x(0;P^*),\ldots,x(N;P^*) \right] = \overline{\mathcal Z}^N\!\left[x^*(0;P^*),\ldots,x^*(N;P^*)\right] \ \overline{\mathcal Z}^{*N}\!\left[x^*(0;P),\ldots,x^*(N;P)\right],
\end{equation}
The product of the partition functions is invariant under the simultaneous exchange
\begin{equation}
    x(w;P) \leftrightarrow x^*(w;P^*)\quad \text{and}\quad x(w;P^*) \leftrightarrow x^*(w;P)\quad \forall w=0,\ldots,N,
\end{equation}
and so the combined system is self-dual. For such systems, the critical manifold is the fixed point of the duality. However, this cannot be imposed in general over all $w$.
\textcite{Takeda2005} conjectured that the principal Boltzmann factors alone determine criticality, giving the condition
\begin{equation}
    \frac{x(0;P_c)}{x^*(0;P_c)}= \frac{x^*(0;P_c^*)}{x(0;P_c^*)}.
    \label{eq:fixed-point}
\end{equation}
(This conjecture is exact for clean (non-disordered) models and for $N=1,2$ replicas on the Nishimori line.) 
Taking $w=w^*=0$ in Eqs.~\eqref{eq:Boltz-x} and \eqref{eq:Boltz-x-dual} gives
\begin{equation}
    x(0;P) = \Ex_{p \sim P} \left[ (1-p)^{N+1} + p^{N+1} \right],\quad x^*(0;P) = 2^{-N/2}.
\end{equation}
To complete the replica trick, we take the quenched limit $N\to0^+$ of $\tfrac1N\ln\!\big[x(0;P_c)/x^*(0;P_c)\big]$. At $N=0$ this ratio equals $1$, so we consider the leading terms in $N$,
\begin{equation}
    \ln\left(2^{N/2} \Ex_{p\sim P} \left[(1-p)^{N+1} + p^{N+1}\right]\right)
    = \frac{N}2\ln2 + \Ex_{p\sim P} \left[ (1-p)\ln(1-p)+ p\ln p \right]N + O(N^2).
\end{equation}
Hence,
\begin{equation}
    \lim_{N\to0^+}\frac1N\ln\frac{x(0;P_c)}{x^*(0;P_c)} = \ln 2\,\Big(\tfrac12-\overline H[P_c]\Big),
\end{equation}
where, with the convention $\log\equiv\log_2$,
\begin{equation}
    \overline H[P] = \Ex_{p\sim P}\left[ -p\log p - (1-p)\log(1-p) \right]
\end{equation}
is the average binary entropy. The fixed-point condition Eq.~\eqref{eq:fixed-point} then gives the desired equation,
\begin{equation}
    \overline H[P_c]+\overline H[P_c^*]=1.
\end{equation}
This derivation relied on integrability, and the validity of exchanging the limit and the expectation. This is justified for $N\geq 0$, as $p^{N+1}\ln p$ is finite for $p\in[0,1]$ and $N\geq 0$.

\subsection{Correlated error parameters}
\label{app:entropy-corr}

We now relax the i.i.d.\ assumption, and let $\bm p$ be drawn from a joint distribution $\mathcal P(\bm p)$ whose single-mechanism marginals are all $P(p)$. The faults remain independent conditional on $\bm p$ [Eq.~\eqref{eq:fault-prob}], but the expectation over $\bm p$ can no longer be moved through the product over mechanisms:
\begin{equation}
    \Ex_{\bm p\sim\mathcal P}\left[\left\langle \widetilde{\mathcal Z}[\bm f;\bm p]^N \right\rangle_{\bm f}\right] = \sum_{\{\bm S^{(r)}\}_{r=1}^N} \Ex_{\bm p\sim\mathcal P}\bigg[\prod_{i=1}^n y(w_i;p_i)\bigg],
\end{equation}
where $y(w;p)$ is the bracketed term of Eq.~\eqref{eq:Boltz-x}, so that $x(w;P) = \Ex_{p\sim P}[y(w;p)]$. The averaged model does not factorize and now has effective interactions between mechanisms; and the single-interaction Boltzmann factors of Eq.~\eqref{eq:Boltz-x} are no longer valid. 
The $\mathbb Z_2^N$ Fourier transform acts on each interaction separately at fixed $\bm p$, so the duality relation of \textcite{Takeda2005} holds for each realization of $\bm p$, with interaction-dependent Boltzmann factors $y(w;p_i)$ and $y^*(w^*;p_i)$ [the bracketed term of Eq.~\eqref{eq:Boltz-x-dual}], and hence, by linearity, after averaging over any $\mathcal P$. Additionally, the Nishimori gauge symmetry requires that the faults be distributed according to Eq.~\eqref{eq:fault-prob} given $\bm p$, which is unchanged.

In place of the principal Boltzmann factor of a single interaction, we therefore use that of the whole system: the weight of the configuration in which every interaction is satisfied in every replica ($w_i=0$ for all $i$),
\begin{equation}
    X_0[\mathcal P] := \Ex_{\bm p\sim\mathcal P}\bigg[\prod_{i=1}^n y(0;p_i)\bigg], \qquad X_0^*[\mathcal P] := \Ex_{\bm p\sim\mathcal P}\bigg[\prod_{i=1}^n y^*(0;p_i)\bigg],
\end{equation}
and impose the fixed-point condition per interaction,
\begin{equation}
    \frac1n\ln\frac{X_0[\mathcal P_c]}{X_0^*[\mathcal P_c]} = \frac1n\ln\frac{X_0^*[\mathcal P_c^*]}{X_0[\mathcal P_c^*]}.
    \label{eq:fixed-point-corr}
\end{equation}
For i.i.d.\ error parameters, $X_0[P^{\otimes n}] = x(0;P)^n$ and $X_0^*[P^{\otimes n}] = x^*(0;P)^n$, and Eq.~\eqref{eq:fixed-point-corr} reduces exactly to Eq.~\eqref{eq:fixed-point}.
At finite $n$, every factor $y(0;p_i)$ equals $1$ at $N=0$, so we again consider the leading terms as $N\rightarrow0^+$,
\begin{equation}
    \prod_{i=1}^n y(0;p_i) = 1 + N\sum_{i=1}^n \left[\frac{\partial y(0;p_i)}{\partial N} \right]_{N=0} + O(N^2), 
    \label{eq:prod_y_N}
\end{equation}
using the product rule for differentiation. For the dual model, we have $X_0^*[\mathcal P_c] = 2^{-nN/2}$. Taking the expectation, logarithm, and limit, we arrive at the familiar expression
\begin{align}
    \lim_{N\to0}\frac1{nN}\ln \frac{X_0[\mathcal P]}{X_0^*[\mathcal P]} &= \frac1n \sum_{i=1}^n\Ex_{\bm p \sim \mathcal P}\left[p_i \ln p_i + (1-p_i)\ln(1-p_i)\right] + \frac12\ln 2 \\ 
    &= \ln 2 \left(\frac12 - \overline H[P]\right),
\end{align}
which depends on $\mathcal P$ only through its marginal $P$. Correlations between mechanisms enter only through covariances between terms at $O(N^2)$ in Eq.~\eqref{eq:prod_y_N}, and so drop out in the quenched limit, exactly as the higher cumulants of $P$ do in the i.i.d.\ case. The remainder of the calculation is unchanged, giving $\overline H[P_c] + \overline H[P_c^*] = 1$ for correlated error parameters with marginals $P_c$ and $P_c^*$. 

Here we take $N\to0$ at finite $n$, before the thermodynamic limit, as required for the quenched average $\Ex[\ln\widetilde{\mathcal Z}] = (\partial\Ex[\widetilde{\mathcal Z}^N])/\partial N|_{N=0}$. Taking $N\to0$ first captures typical disorder configurations, but not rare regions. If $\mathcal P$ allows correlated regions of noisy error parameters large enough to support a logical failure, with a probability that does not vanish as $d\to\infty$, these regions can dominate the logical failure rate and remove the threshold even when the condition on the marginals is satisfied~\cite{Sriram2025}. We therefore require such regions to be exponentially suppressed in $d$, as holds when $\mathcal P$ has a finite correlation length along at least one direction spanned by every logical operator. Error parameters fixed in time satisfy this, since a logical failure then requires a correlated region extending across the code in space, which occurs with probability exponentially small in $d$. Smaller rare regions can still produce Griffiths effects, which modify the approach to asymptotic scaling near threshold~\cite{Sriram2025}.

\subsection{Mechanisms with more than two outcomes}
\label{app:entropy-corr-noise}

We now allow a mechanism to produce more than two outcomes, as for the depolarizing channel of Eq.~\eqref{eq:noise-channel-dep}. Recall that we can describe a mechanism as producing faults $f_i$ in some abelian group $G$. For example, the $m$-qubit Paulis modulo phase, $G \cong \mathbb Z_2^{2m}$. We write $\PP[f_i;p_i]$ for each fault probability; for depolarizing noise, $\PP[0;p] = 1-p$ and $\PP[f;p] = p/(|G|-1)$ for $f\neq0$. We take all mechanisms to be of the same type, and keep the joint parameter distribution $\mathcal P$ of Appendix~\ref{app:entropy-corr} (with identical marginals $P$).

The statistical-mechanical models remain Ising-like, with spins $S \in \{-1,1\}$~\cite{Chubb2021,Bombin2012}. Writing $(\bm g_k)_i \in G$ for the action of the generator $\bm g_k$ at mechanism $i$, a spin configuration applies $g_i(\bm S) := \sum_{k:\,S_k=-1}(\bm g_k)_i$ at mechanism $i$, and the Boltzmann weight is (for normalized Hamiltonian $\widetilde{\mathcal H}$)
\begin{equation}
    e^{-\widetilde{\mathcal H}[\bm S,\bm f;\bm p]} = \prod_{i=1}^n \PP\big[f_i + g_i(\bm S);p_i\big].
\end{equation}
For $G = \mathbb Z_2$ this is the Hamiltonian of Sec.~\ref{sec:bg-sm} up to the $\bm S$-independent normalization factor $\prod_i[p_i(1-p_i)]^{1/2}$. For single-qubit depolarizing noise, $g_i(\bm S)$ is fixed by two Ising spins (its $X$ and $Z$ components); the $Y$ faults couple the terms together, giving four-spin interactions~\cite{Bombin2012}.

Averaging over $\bm f$ at fixed $\bm p$ as before, the replicated partition function is
\begin{equation}
    \Ex_{\bm p\sim\mathcal P}\left[\left\langle \widetilde{\mathcal Z}[\bm f;\bm p]^N \right\rangle_{\bm f}\right] = \sum_{\{\bm S^{(r)}\}_{r=1}^N} \Ex_{\bm p\sim\mathcal P}\bigg[\prod_{i=1}^n y(\vec g_i;p_i)\bigg], \qquad y(\vec g;p) := \sum_{f\in G}\PP[f;p]\prod_{r=1}^N\PP[f+g_r;p],
\end{equation}
where $\vec g_i = \big(g_i(\bm S^{(1)}),\ldots,g_i(\bm S^{(N)})\big) \in G^N$. For $G = \mathbb Z_2$, $y(\vec g;p) = y(w;p)$ is the normalized binary weight, with $w$ the number of nonzero entries of $\vec g$. The duality of \textcite{Takeda2005} carries over with the $\mathbb Z_2^N$ Fourier transform of each interaction replaced by the $G^N$ Fourier transform~\cite{Bombin2012,Ohzeki2012}; as in Appendix~\ref{app:entropy-corr}, it acts on each interaction at fixed $\bm p$. The principal Boltzmann factors are
\begin{align}
    y(\vec 0;p) &= \sum_{f\in G}\PP[f;p]^{N+1},\\
    y^*(\vec 0;p) &= |G|^{-N/2}\sum_{\vec g\in G^N} y(\vec g;p) = |G|^{-N/2},
\end{align}
using $\sum_{g\in G}\PP[f+g;p] = 1$. Defining $X_0[\mathcal P]$ and $X_0^*[\mathcal P]$ from these factors as in Appendix~\ref{app:entropy-corr}, the quenched limit gives
\begin{equation}
    \lim_{N\to0}\frac1{nN}\ln\frac{X_0[\mathcal P]}{X_0^*[\mathcal P]} = \Ex_{p\sim P}\bigg[\tfrac12\ln|G| + \sum_{f\in G}\PP[f;p]\ln\PP[f;p]\bigg] = \ln2\,\Big(\tfrac12\log|G| - \overline H_{G}[P]\Big),
\end{equation}
where
\begin{equation}
    \overline H_{G}[P] := \Ex_{p\sim P}\bigg[-\sum_{f\in G}\PP[f;p]\log\PP[f;p]\bigg]
\end{equation}
is the average Shannon entropy of a mechanism's fault distribution, with $\overline H_{\mathbb Z_2}[P] = \overline H[P]$, the average binary entropy. Since $G$ is isomorphic to its Fourier dual, the dual model has mechanisms of the same type, and the fixed-point condition, Eq.~\eqref{eq:fixed-point-corr}, gives
\begin{equation}
    \overline H_{G}[P_c] + \overline H_{G}[P_c^*] = \log|G|.
\end{equation}
This is analogous to Eq.~\eqref{eq:critical-H-group}, with $\overline H_G[P_c] = \tfrac12\log|G|$ for a self-dual model. The rare-region condition of Appendix~\ref{app:entropy-corr} remains. The result also relies on the $G^N$ Fourier transform mapping the model onto its dual, which holds for the toric code under depolarizing noise~\cite{Bombin2012}.

For depolarizing mechanisms, the grouping property of the entropy gives $-\sum_f\PP[f;p]\log\PP[f;p] = H(p) + p\log(|G|-1)$, so that
\begin{equation}
    \overline H_G[P] = \overline H[P] + \overline p\log(|G|-1).
    \label{eq:H-depolarizing}
\end{equation}
For the toric code under homogeneous single-qubit depolarizing noise ($|G| = 4$, self-dual), $H(p_c) + p_c\log3 = 1$ gives $p_c = 0.18929$, reproducing the hashing bound~\cite{Bennett1996} and the duality estimate of Ref.~\cite{Ohzeki2012}, compared with $p_c = 0.189(3)$ from Monte Carlo simulations~\cite{Bombin2012}. Qubit loss at rate $q$ corresponds to the bimodal distribution $P(p) = (1-q)\,\delta(p-p_\mathrm{dep}) + q\,\delta(p-\tfrac34)$, since a lost qubit is fully depolarized, with entropy $\log4 = 2$. The condition $2q + (1-q)\big[H(p_c) + p_c\log3\big] = 1$ then reproduces the single-cluster estimates of Ref.~\cite{Ohzeki2012} for all loss rates, the depolarizing analogue of Appendix~\ref{app:erasure}.

\subsection{Mischaracterized entropy thresholds}
\label{app:entropy-off-nish}

Here we extend the calculation of Appendix~\ref{app:entropy} to the mischaracterized decoders of Sec.~\ref{sec:mis-near}, deriving Eq.~\eqref{eq:critical-mischar}. The spin model now has couplings $K(\hat p_i)$ set by the assumed parameters, while the reference fault $\bm f$ is still drawn from $\PP[\bm f;\bm p]$ with the true parameters. We assume that the pairs $(p_i,\hat p_i)$ are drawn i.i.d.\ from the joint parameter distribution $\Pi$.

Following Appendix~\ref{app:entropy} with $K(p_i) \mapsto K(\hat p_i)$, the $N$-th moment of the normalized partition function is
\begin{equation}
    \Ex_{(\bm p,\hat{\bm p}) \sim \Pi}\left[\left\langle \widetilde{\mathcal Z}[\bm f;\hat{\bm p}]^N \right\rangle_{\bm f}\right] = \sum_{\{\bm S^{(r)}\}_{r=1}^N} \prod_{i=1}^n x(w_i;\Pi) := \overline{\mathcal Z}^N[x(0;\Pi),\ldots,x(N;\Pi)],
\end{equation}
where $\langle\cdot\rangle_{\bm f}$ averages over $\bm f \sim \PP[\bm f;\bm p]$, $w_i$ is defined as before, and the averaged Boltzmann factors are
\begin{equation}
    x(w;\Pi) = \Ex_{(p,\hat p)\sim \Pi}\left[(1-p)(1-\hat p)^{N-w}\hat p ^w + p\hat p^{N-w}(1-\hat p)^w\right].
    \label{eq:Boltz-x-mis}
\end{equation}
The $\mathbb Z_2^N$ Fourier transform acts on each interaction separately and does not use the Nishimori relation, so the dual Boltzmann factors follow from Eq.~\eqref{eq:Boltz-x-dual} by the same replacement,
\begin{equation}
    x^*(w^*;\Pi) = \Ex_{(p,\hat p)\sim \Pi}\bigg[2^{-N/2}(1-2\hat p)^{w^*} \left([1-p]+(-1)^{w^*}p\right)\bigg].
    \label{eq:Boltz-x-dual-mis}
\end{equation}
The duality relation of \textcite{Takeda2005} therefore holds unchanged with $P \to \Pi$ and $P^* \to \Pi^*$, where $\Pi^*$ is the joint parameter distribution of the dual model.

We now impose the fixed-point condition of Eq.~\eqref{eq:fixed-point} on the principal Boltzmann factors,
\begin{equation}
    \frac{x(0;\Pi_c)}{x^*(0;\Pi_c)}= \frac{x^*(0;\Pi_c^*)}{x(0;\Pi_c^*)}.
    \label{eq:fixed-point-mis}
\end{equation}
These are given by
\begin{equation}
    x(0;\Pi) = \Ex_{(p,\hat p)\sim\Pi} \left[(1-p)(1-\hat p)^N + p\hat p^N\right],\quad x^*(0;\Pi) = 2^{-N/2}.
\end{equation}
If $\Ex[p \ln \hat p]$ and $\Ex[(1-p)\ln(1-\hat p)]$ are finite, we can differentiate inside the expectation at $N=0^+$. The quenched limit is then
\begin{align}
    \lim_{N\to0}\frac1N\ln\frac{x(0;\Pi_c)}{x^*(0;\Pi_c)}
    &= \Ex_{(p,\hat p)\sim\Pi_c}\!\big[p\ln \hat p+(1-p)\ln(1-\hat p)\big]+\tfrac12\ln2 \\
    &= \ln 2\,\Big(\tfrac12-\overline H[\Pi_c]\Big),
\end{align}
with $\overline H[\Pi]$ the average binary cross entropy, Eq.~\eqref{eq:bce}. As in the characterized cases, only the first derivative enters the replica limit. The fixed-point condition Eq.~\eqref{eq:fixed-point-mis} then gives
\begin{equation}
    \overline H[\Pi_c]+\overline H[\Pi_c^*]=1.
\end{equation}
Exchanging the limit and the expectation requires $\Ex_\Pi|p\ln\hat p| < \infty$ and $\Ex_\Pi|(1-p)\ln(1-\hat p)| < \infty$, which does not hold for generic mischaracterization $\Pi$.
Given the replica limit, this derivation is exact except for the fixed-point condition, Eq.~\eqref{eq:fixed-point-mis}. On the Nishimori line this conjecture is supported by symmetry and the multicritical point lying on the line~\cite{Nishimori2002,Takeda2005}; off the line neither argument applies. The argument of Appendix~\ref{app:entropy-corr} applies unchanged when the pairs $(p_i,\hat p_i)$ are correlated across mechanisms with pair marginal $\Pi$, since the $N$-derivatives above are again linear in the joint distribution, subject to the same rare-region condition.

\section{Erasure bit-flip channels}
\label{app:erasure}

For self-dual models with duality-constrained critical entropy $H_c = \frac12$, we can construct a parameter distribution that includes both erasures (with probability $q$) and single-qubit independent $X$ noise (with probability $p_X$), by
\begin{equation}
    P(p) = (1-q)\delta(p-p_X) + q\delta(p-\tfrac12).
\end{equation}
Its average binary entropy is 
\begin{equation}
    \overline H[P] = (1-q)H(p_X) + qH(\tfrac12) = q - (1-q)\left[p_X\log p_X + (1-p_X)\log(1-p_X)\right].
\end{equation}
Equating this to $\frac12$ along the critical contour gives the analytic expression for the critical erasure rate at a given $p_X$:
\begin{equation}
    q_c = \frac{\frac12 + p_X\log p_X + (1-p_X)\log(1-p_X)}{1 + p_X\log p_X + (1-p_X)\log(1-p_X)}.
\end{equation}
This agrees with numeric and other analytical predictions of the erasure--bit-flip critical phase boundary~\cite{Ohzeki2012, English2026}.

\section{Numerical details and supplementary results}
\label{app:sup-numerics}

In this section, we provide additional details for the numerical simulations reported in this work, and showcase supplementary results.

\begin{figure}[t]
    \centering
\includegraphics[width=\linewidth]{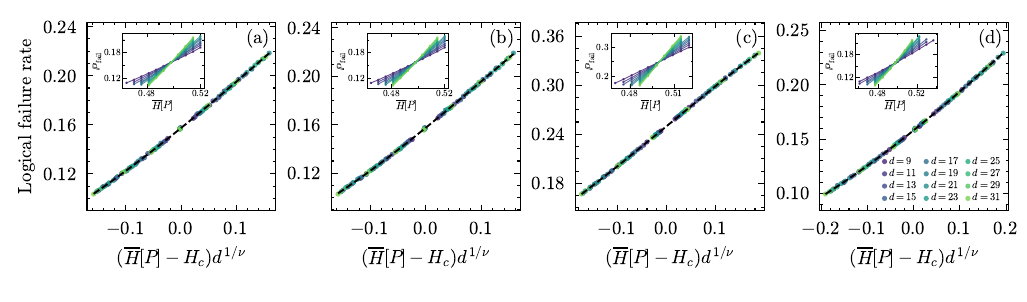}
    \caption{Data collapse of the logical failure rate of $d=9,\ldots,31$ surface codes under ML decoding, against code-capacity bit-flip noise, with error parameters drawn from
        (a) log-normal
        (b) Gaussian
        (c) erasure
        (d) homogeneous distributions.
        \textit{Inset}: the data prior to rescaling. Error bars are smaller than data points, so are omitted.
        }
    \label{fig:code_capacity_data_collapse}
\end{figure}

We obtain the data in Fig.~\ref{fig:intro-thresholds} using $N=10^{6}$ samples, across code distances of length $d=9,\ldots,31$, using optimal ML decoding performed with \textit{Planar}~\cite{Cao2025}. In Fig.~\ref{fig:code_capacity_data_collapse}, we show data collapses of the logical failure rate from this data against the scaling variable $x=(\overline{H}[P]-H_{c})d^{1/\nu}$ for planar surface codes with bit-flip code-capacity noise drawn from a variety of distributions. 

We estimate the critical mean entropy $H_{c}$ by fitting a cubic polynomial in the scaling variable. The fit results are shown in Table~\ref{tab:H_c-and-nu}. Parentheses denote two standard errors in the statistical uncertainties obtained from the fit. These uncertainties do not include systematic effects associated with finite lattice sizes or the scaling ansatz. Residual dependence on the minimum included distance was observed. Therefore, the quoted errors should not be interpreted as the full uncertainty in the asymptotic critical parameters. Nonuniversal additive corrections ($cd^{-\omega}$~\cite{Wang2003}) to the third-order polynomial ansatz did not improve the fit. Nonetheless, across the four distribution types, we observe the critical $H_{c}$ lies in a small window centered at $H_{c}\simeq 0.5$. 

\begin{table}[t]
    \centering
    \caption{Fitted critical average entropy and critical exponent for optimal ML decoding of planar surface codes in the bit-flip code-capacity setting.}
    \label{tab:H_c-and-nu}
    \setlength{\tabcolsep}{8pt}
    \begin{tabular}{lcc}
    \toprule
    $P(p)$ distribution & $H_{c}$ & $\nu$ \\
    \midrule
         Log-normal & \(0.4980(2)\) & \(1.54(2)\) \\
         Gaussian & \(0.4982(2)\) & 	\(1.54(2)\) \\
         Erasure & \(0.50002(8)\) & \(1.34(1)\) \\
         Homogeneous & \(0.4971(2)\) & \(1.53(2)\) \\
    \bottomrule
    \end{tabular}
\end{table}

Erasure noise was consistent within two standard errors with the values predicted by bond percolation on a square lattice~\cite{Stace2009}. The critical exponents $\nu$ for log-normal, Gaussian and homogeneous noise agreed up to uncertainty (at two standard errors), which may be consistent with shared universality across these distributions~\cite{Picco2006,Wan2026}.

In Fig.~\ref{fig:intro-decay-rate}, we fit the decay rates from linear regression on $-\ln \mathbb{P}_{\mathrm{fail}}$ against $d$. In the deep subthreshold regime (for $\overline{p}<5\times10^{-3}$), we use importance sampling with $d=13,\ldots, 19$ and $N=10^{8}$ samples. This data is also presented in Fig.~\ref{fig:alpha-codecapacity-bhat}. For larger $\overline{p}$, we use direct sampling for $d=5,\ldots,17$ with $N=10^{6}$ samples. We use the ansatz
\[
\alpha(\overline{B}[P])=-\ln \overline{B}[P]-2\overline{B}[P]-c\overline{B}[P]^{2}
\]
obtaining $c=-1.91$. The constant and linear terms follow the lower bound derived in Sec.~\ref{sec:char-sub-path}; the fitted quadratic term is an empirical correction for the code distances and noise families simulated.

We show the number of distances and number of samples per data point simulated in the remaining figures of the main text in Table~\ref{tab:fig_d_N}. 
We also show the parameters of the log-normal distributions for the results presented in Fig.~\ref{fig:alpha_against_KL} in Table~\ref{tab:alpha_against_KL}.

\begin{table}[t]
    \centering
    \caption{Simulation details for figures in the main text.}
    \label{tab:fig_d_N}
    \setlength{\tabcolsep}{8pt}
    \begin{tabular}{ccc}
    \toprule
    Figure & Distances $(d)$ & Samples per data point $(N)$ \\
    \midrule
    \ref{fig:intro-thresholds}/\ref{fig:code_capacity_data_collapse} & $9,11,\ldots,31$ & $10^6$ \\
    \ref{fig:char_contour} & $11,13$ & $10^{7}$\\
    \ref{fig:LER_vsHbar_bitflip_codecapacity} & $9,11,\ldots,31$ & $10^{6}$\\
    \ref{fig:alpha-codecapacity-bhat} & $13,15,\ldots,19$ & $10^{8}$\\
    \ref{fig:LER_vsHbar_phenomenological_MWPM} & $5,7,\ldots,17$ & $10^{6}$\\
    \ref{fig:alpha-phen-bhat} & $15,17,\ldots,21$ & $10^{8}$\\
    \ref{fig:LER_vsHbar_phenomenological_MWPM_fixed_time} & $5,7,\ldots,17$ & $10^{6}$\\
    \ref{fig:mischar_threshold_code_cap} & $9,11,\ldots,31$ & $10^{5}$\\
    \ref{fig:alpha_against_KL} & $5,7,\ldots,13$ & $10^{5}-10^{6}$\\
    \ref{fig:mischar_grid} & $13,15,\ldots,19$ & $10^{8}$ \\
    \bottomrule
    \end{tabular}
\end{table}

\begin{table}[t]
    \centering
    \caption{Log-normal parameters used in the simulations for results presented in Fig.~\ref{fig:alpha_against_KL}.}
    \label{tab:alpha_against_KL}
    \setlength{\tabcolsep}{8pt}
    \begin{tabular}{cccc}
    \toprule
    $\overline{p}$ & $\sigma$ & $\overline{H}[P]$ & $\Delta H$\\
    \midrule
        0.01 & 1.6593 & 0.0628 & 0.018\\
        0.02 & 1.1150 & 0.1234 & 0.018\\
        0.03 & 0.8984 & 0.1764 & 0.018\\
        0.04 & 0.7719 & 0.2243 & 0.018\\
        0.05 & 0.686 & 0.2684 & 0.018\\
        0.06 & 0.6226 & 0.3094 & 0.018\\
        0.07 & 0.5732 & 0.3479 & 0.018\\
        0.08 & 0.5332 & 0.3842 & 0.018\\
        0.09 & 0.5 & 0.4185 & 0.018\\
    \bottomrule
    \end{tabular}
\end{table}

\subsection{Circuit-level noise simulations}\label{app:CLN_sims}

\begin{figure}
    \centering
\begin{tikzpicture}
    \node at (0,0) {\includegraphics[width=0.8\linewidth]{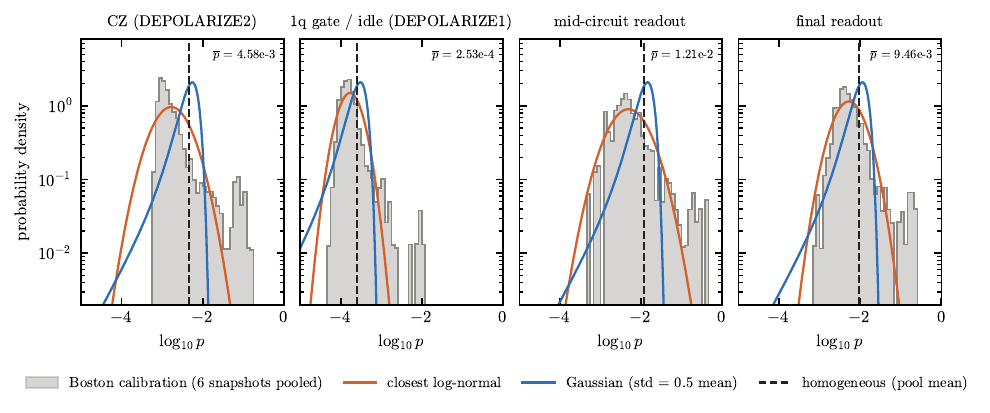}};
    \node at (0,-6.5) {\includegraphics[width=0.8\linewidth]{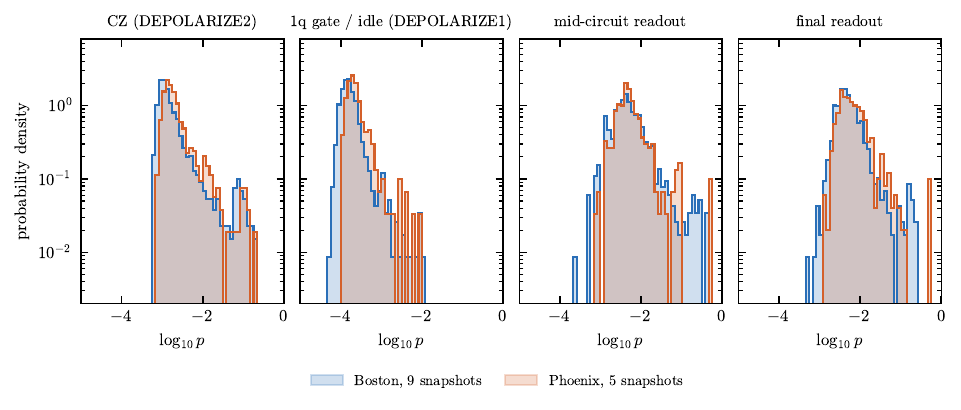}};
    \node at (-6.5,2.8) {(a)};
    \node at (-6.5,-3.7) {(b)};
\end{tikzpicture}
    \caption{(a) Distributions obtained from the device \texttt{ibm\_boston} and fitted log-normal and Gaussian distributions superimposed. The histogram (gray) shows the probability density of noise parameter $p$ obtained from six days' worth of calibration data from the device, converted to a Pauli noise model. The four error mechanisms resulting from $CZ$ gates, single-qubit gates/idling, and mid-circuit and final readout are shown on the four panels. We also show the fits for log-normal and Gaussian distributions used for simulations, and additionally plot the mean of the calibration noise data, used for the ``homogeneous noise'' simulations. (b) Distributions obtained from nine \texttt{ibm\_boston} snapshots compared to five \texttt{ibm\_phoenix} snapshots. The features of the $CZ$ and single-qubit noise distributions appear very similar between both devices, in particular the heavy tails at high-$p$ in the former. The distributions are also fairly consistent for readout errors, although the Boston distributions have somewhat heavier tails at high-$p$.}
    \label{fig:error_rate_hists}
\end{figure}

Here we provide further details on the circuit-level noise simulations used to obtain the results in Fig.~\ref{fig:circuit_level_data}. We obtained multiple days' worth of calibration data for superconducting devices \texttt{ibm\_boston} and \texttt{ibm\_phoenix} from Ref.~\cite{ibm_compute}. We then pool all days' data for each device and obtain distributions over Pauli noise channels based on this data. We map IBM-reported noise information to Pauli noise channels in the following way~\cite{mclauchlan2026errorcorrectedmemorylogicheavyhex}:
\begin{itemize}
    \item For each coupler in the device, we let the reported $CZ$ gate error rate equal the error rate $p$ for a 2-qubit unbiased depolarizing noise channel (``DEPOLARIZE2''), occurring after CNOT gates in the syndrome extraction circuit. 
    \item For each qubit in the device, we let the reported idle error (``ID error'') equal the error rate $p$ for a single-qubit unbiased depolarizing noise channel (``DEPOLARIZE1''), occurring after single-qubit Clifford gates in the syndrome extraction circuit and during idling operations. This reported error is equal to all other single-qubit operation errors.
    \item We let the reported ``MEASURE\_2'' error be the classical readout error during mid-circuit measurements.
    \item We let the reported ``MEASURE'' error be the classical readout error for end-of-circuit measurements. 
\end{itemize}
This results in the histograms of error rates for 2-qubit, single-qubit and measurement errors shown in Fig.~\ref{fig:error_rate_hists}. We call these the device noise distributions.

To obtain the data in Fig.~\ref{fig:circuit_level_data}, we produce scaled device distributions $P_\lambda(p) = \lambda^{-1} P\!\left(p\lambda^{-1}\right)$ for $\lambda\geq 1$ and $P(p)$ the device noise distribution. That is, we scale the sampled noise by $\lambda$, which shifts the mean of $\ln p$ by $\ln \lambda$ and leaves its standard deviation unchanged. For each $\lambda$ and $d$, we perform $150$ sets of noise samples, each setting the noise for an entire memory experiment, and then perform $2000$ simulation shots ($10^4$ for lower values of $\lambda$) for each sample. 

We then produce log-normal and Gaussian fits to the device noise distributions, shown in Fig.~\ref{fig:error_rate_hists}. The log-normal distributions were set such that the mean and standard deviation of $\ln p$ were equal to those of the respective device noise distributions. Meanwhile, the Gaussian distributions were chosen to have the same mean and half the standard deviations of the respective device noise distributions. For the homogeneous distributions we set all noise parameters to $\bar p$, the mean of the respective $P(p)$. We perform $3\times 10^4$ samples in total for these three types of distribution, split between error-correction shots and noise samples.

For all simulations, we run $Z$ memory experiments for the rotated surface code using Stim~\cite{Gidney2021} and decode using PyMatching~\cite{Higgott2025sparseblossom}. We do not apply noise to resets in these circuits, since \texttt{ibm\_boston} does not have the capacity for mid-circuit resets so we do not have complete data for this type of noise.

\end{document}